\pdfoutput=1
\documentclass[a4paper,fleqn,usenatbib,useAMS]{mnras}
\usepackage{color}
\usepackage{aas_macros}
\usepackage{adjustbox}
\usepackage{todonotes}
\usepackage{bm} 
\usepackage{times}

\usepackage{amsmath}
\usepackage{cleveref}

\newcommand{\ims}{\texttt{im3shape}}
\newcommand{\ngm}{\texttt{ngmix}}

\newcommand{\bea}{\begin{eqnarray}}
\newcommand{\be}{\begin{equation}}
\newcommand{\ben}{\begin{enumerate}}
\newcommand{\bi}{\begin{itemize}}
\newcommand{\eea}{\end{eqnarray}}
\newcommand{\ee}{\end{equation}}
\newcommand{\ei}{\end{itemize}}
\newcommand{\een}{\end{enumerate}}

\newcommand{\om}{\Omega_\mr m}

\newcommand{\mr}{\mathrm}

\newcommand{\kkE}{\kappa_{\rm CMB}\gamma_{\rm E}}
\newcommand{\kkB}{\kappa_{\rm CMB}\gamma_{\rm B}}
\newcommand{\kkI}{\kappa_{\rm CMB}\gamma_{\rm I}}

\usepackage{amsmath,amssymb}

\def\eprinttmp@#1arXiv:#2 [#3]#4@{
\ifthenelse{\equal{#3}{x}}{\href{http://arxiv.org/abs/#1}{#1}
}{\href{http://arxiv.org/abs/#2}{arXiv:#2} [#3]}}

\providecommand{\eprint}[1]{\eprinttmp@#1arXiv: [x]@}
\newcommand{\adsurl}[1]{\href{#1}{ADS}}

\title[Cross-correlation of gravitational lensing from DES SV data with SPT \& Planck lensing]{Cross-correlation of gravitational lensing from DES Science Verification data with SPT and \emph{Planck} lensing}

\author[]{
\parbox{\textwidth}{
\Large
D.~Kirk$^{1}$,\thanks{Corresponding author: \texttt{drgk@star.ucl.ac.uk}}
Y.~Omori$^{2}$,\thanks{Corresponding author: \texttt{yomori@physics.mcgill.ca}}
A.~Benoit-L{\'e}vy$^{3,4,5,1}$,
R.~Cawthon$^{5,6}$,
C.~Chang$^{7}$,
P.~Larsen$^{8,9}$,
A.~Amara$^{7}$,
D.~Bacon$^{10}$,
T.~M.~Crawford$^{5,6}$,
S.~Dodelson$^{4,5,11}$,
P.~Fosalba$^{12}$,
T.~Giannantonio$^{8,9,13}$,
G.~Holder$^{2}$,
B.~Jain$^{14}$,
T.~Kacprzak$^{7}$,
O.~Lahav$^{1}$,
N.~MacCrann$^{15}$, 
A.~Nicola$^{7}$,
A.~Refregier$^{7}$,
E.~Sheldon$^{16}$,
K.~T.~Story$^{5,11}$,
M.~A.~Troxel$^{15}$,
J.~D.~Vieira$^{17}$,
V.~Vikram$^{18}$,
J.~Zuntz$^{15}$,
T. M. C.~Abbott$^{19}$,
F.~B.~Abdalla$^{1,20}$,
M.~R.~Becker$^{16,21}$,
B.~A.~Benson$^{4,6,5}$,
G.~M.~Bernstein$^{14}$,
R.~A.~Bernstein$^{22}$,
L.~E.~Bleem$^{18,5}$,
C.~Bonnett$^{23}$,
S.~L.~Bridle$^{15}$,
D.~Brooks$^{1}$,
E.~Buckley-Geer$^{4}$,
D.~L.~Burke$^{21,24}$,
D.~Capozzi$^{10}$,
J.~E.~Carlstrom$^{6,5,11}$,
A.~Carnero~Rosell$^{25,26}$,
M.~Carrasco~Kind$^{17,27}$,
J.~Carretero$^{12,23}$,
M.~Crocce$^{12}$,
C.~E.~Cunha$^{21}$,
C.~B.~D'Andrea$^{10,28}$,
L.~N.~da Costa$^{25,26}$,
S.~Desai$^{29,30}$,
H.~T.~Diehl$^{4}$,
J.~P.~Dietrich$^{30,29}$,
P.~Doel$^{1}$,
T.~F.~Eifler$^{14,31}$,
A.~E.~Evrard$^{32,33}$,
B.~Flaugher$^{4}$,
J.~Frieman$^{5,4}$,
D.~W.~Gerdes$^{33}$,
D.~A.~Goldstein$^{34,35}$,
D.~Gruen$^{21,36,24,30}$,
R.~A.~Gruendl$^{17,27}$,
K.~Honscheid$^{37,38}$,
D.~J.~James$^{19}$,
M.~Jarvis$^{14}$,
S.~Kent$^{4}$,
K.~Kuehn$^{39}$,
N.~Kuropatkin$^{4}$,
M.~Lima$^{40,25}$,
M.~March$^{14}$,
P.~Martini$^{37,41}$,
P.~Melchior$^{42}$,
C.~J.~Miller$^{32,33}$,
R.~Miquel$^{43,23}$,
R.~C.~Nichol$^{10}$,
R.~Ogando$^{25,26}$,
A.~A.~Plazas$^{31}$,
C.~L.~Reichardt$^{44}$,
A.~Roodman$^{21,24}$,
E.~Rozo$^{45}$,
E.~S.~Rykoff$^{21,24}$,
M.~Sako$^{14}$,
E.~Sanchez$^{46}$,
V.~Scarpine$^{4}$,
M.~Schubnell$^{33}$,
I.~Sevilla-Noarbe$^{46}$,
G.~Simard$^{2}$,
R.~C.~Smith$^{19}$,
M.~Soares-Santos$^{4}$,
F.~Sobreira$^{47,25}$,
E.~Suchyta$^{14}$,
M.~E.~C.~Swanson$^{27}$,
G.~Tarle$^{32}$,
D.~Thomas$^{10}$,
R.~H.~Wechsler$^{16,21,24}$,
J.~Weller$^{29,30,36}$
\emph{(Affiliations are listed at the end of paper)} 
}
}

\begin{document}

\date{}

\pagerange{\pageref{firstpage}--\pageref{lastpage}} 
\pubyear{2015}

\maketitle

\label{firstpage}

\begin{abstract}

We measure the cross-correlation between weak lensing of galaxy images and of the cosmic microwave background (CMB) on the same patch of sky. The effects of gravitational lensing on different sources will be correlated if the lensing is caused by the same mass fluctuations. We use galaxy shape measurements from 139 deg$^{2}$ of the Dark Energy Survey (DES) Science Verification data and overlapping CMB lensing from the South Pole Telescope (SPT) and \emph{Planck}. The DES source galaxies have a median redshift of $z_{\rm med} {\sim} 0.7$, while the CMB lensing kernel is broad and peaks at $z{\sim}2$. The resulting cross-correlation is maximally sensitive to mass fluctuations at $z{\sim}0.44$. Assuming the \emph{Planck} 2015 best-fit cosmology, the amplitude of the DES$\times$SPT cross-power is found to be $A_{\rm SPT} = 0.88 \pm 0.30$ and that from DES$\times$\emph{Planck} to be $A_{\rm \emph{Planck}} = 0.86 \pm 0.39$, where $A=1$ corresponds to the theoretical prediction and the errors are 68\% confidence limits. These are consistent with the expected signal and correspond to significances of $2.9 \sigma$ and $2.2 \sigma$ respectively. We demonstrate that our results are robust to a number of important systematic effects including the shear measurement method, estimator choice, photometric redshift uncertainty and CMB lensing systematics. Significant intrinsic alignment of galaxy shapes would increase the cross-correlation signal inferred from the data; we calculate a value of $A = 1.08 \pm 0.36$ for DES$\times$SPT when we correct the observations with a simple IA model. With three measurements of this cross-correlation now existing in the literature, there is not yet reliable evidence for any deviation from the expected LCDM level of cross-correlation, given the size of the statistical uncertainties and the significant impact of systematic errors, particularly IAs. Future data sets, including those from upcoming releases of DES and SPT, will cover more sky area and provide both greater depth and higher resolution, making this correlation a potentially very powerful cosmological tool. We provide forecasts for the expected signal-to-noise of the combination of the five-year DES survey and SPT-3G.

\end{abstract}

\begin{keywords}
gravitational lensing: weak; methods: data analysis; cosmic microwave background\vspace{-1.7cm}
\end{keywords}

\newpage
\clearpage

\section{Introduction}\label{sec:intro}
Weak lensing (WL) refers to the coherent bending, by gravity, of light from distant sources as it passes through the large-scale massive structures of the Universe. WL is a powerful tool for studying the distribution and evolution of large-scale structure in the Universe because it is directly sensitive to both dark matter and baryonic matter. 
Here we focus on WL of two background sources: lensing of galaxies (which we will refer to as GWL) and lensing of the cosmic microwave background (referred to as CMBWL). 

Observations of GWL rely on accurately measuring the shapes of a large number of small, faint galaxies, which are often at the edge of our survey magnitude detection limits.
GWL surveys have steadily improved since the first measurements at the turn of the millennium \citep{BRE00,Kaiser:2000if,Wittman:2000tc,van_Waerbeke:2000rm}, most notably with the Canada-France-Hawaii Telescope Lensing Survey \citep[CFHTLenS,][]{heymans2012}, and are now a viable probe of cosmology.
Detection of CMBWL exploits mode couplings in the temperature and polarisation fields of the CMB, which are negligible in absence of WL~\citep{BS87,OH03}. This technique has been used in numerous studies to date \citep{HS03,SZD07,vanengelen12,planck13xvii,das14,planck15xv,story15}. \citet{LC06} provide a comprehensive review of CMBWL. 



GWL and CMBWL signals from the same patch of sky are expected to be correlated as both are, in part, sourced by the gravitational potentials of the same large-scale mass fluctuations. The cross-correlation of the lensing measurements from two such different sources offers a number of important applications. First, it provides a powerful check of systematics for cosmic shear measurements.
For example, \citet{V12} suggested that this cross-correlation can be used to mitigate shear measurement bias, whether from noise or complex galaxy morphologies \citep{jarvisetal2015}, to which CMBWL is insensitive. The same is true of other observational and astrophysical systematics such as modelling the point spread function (PSF) and galaxy intrinsic alignments (IAs), though the estimation of lensing potential from CMB maps is prone to some multiplicative errors of its own. Taking GWL and CMBWL together, there is considerable scope for calibration of these bias terms through cross-correlation of the GWL and CMBWL signals.
In addition, CMB lensing offers an extra high-redshift source bin that can be included in joint analyses of late-universe probes \citep{V13} to study late-time dark energy or modifications to gravity.

Two measurements of the GWL$\times$CMBWL cross-correlation have previously been reported: Atacama Cosmology Telescope (ACT) lensing data crossed with 121 deg$^{2}$ of galaxy lensing convergence as measured by the Canada-France-Hawaii Telescope (CFHT) Stripe 82 Survey \citep{HLD+13}, and \emph{Planck} lensing crossed with 140 deg$^{2}$ galaxy lensing data as measured by CFHTLenS \citep{LH15}. Both report low detected signals compared with expectations for the \emph{Planck} best-fit cosmology, with \citet{LH15} reporting a particularly low signal at roughly half the expected amplitude, a $2\sigma$ discrepancy. In this paper we aim to obtain a measurement of the GWL$\times$CMBWL cross-correlation using new data from DES Science Verification (SV) and CMBWL maps from the South Pole Telescope (SPT) and \emph{Planck}. Our measurement has slightly deeper CMBWL data (SPT compared with ACT and \emph{Planck}), similar sky coverage, but slightly shallower GWL measurements compared to those used in previous GWL$\times$CMBWL results. Importantly, with DES and SPT, we employ different GWL and CMBWL data to those used before for this cross-correlation and cover a different patch of sky to the previous analyses. Our results therefore serve as an independent check on the measurements made by \citet{HLD+13} and \citet{LH15}.

We begin by describing the relevant theory and formalism for GWL and CMBWL in \Cref{sec:theory}. In \Cref{sec:data} we describe our data from DES, SPT and \emph{Planck}. We describe our methods in \Cref{sec:results_GKharmonic_implementation} and present our measurements of the GWL$\times$CMBWL cross-correlation in \Cref{sec:results}. In \Cref{sec:systematics} we demonstrate that our results are robust to a variety of important systematic effects and consistency checks. In \Cref{sec:forecasts} we compare the power of the current data with that expected from the full DES survey and SPT-3G. We discuss the implications of our measurements, their relation to previous results and the future potential of this cross-correlation in \Cref{sec:discussion}. Throughout this paper we employ \emph{Planck} 2015 cosmology (TT+TE+EE+lowP+lensing+ext) with $\Omega_{\rm b}=0.049$, $\Omega_{\rm m}=0.309$, $\Omega_{\Lambda}=0.691$, $\sigma_{8}=0.816$, $h=0.677$.

\section{Theory}
\label{sec:theory}

In this paper we consider two light sources that experience weak lensing: galaxies and the CMB. 
Two particularly useful quantities associated with the distortion of light are the spin-0 convergence field, $\kappa$, and spin-2 shear field, $\gamma$ \citep[see for example][for details and definitions of the lensing quantities]{bartelmann01,MVW+08,HJ08}. Both are derivatives of the lensing potential, which describes the strength of lensing for a given configuration of source, lens and observer.
In GWL, the main observable is shear, which is measured by the distortions of the source galaxy shapes.\footnote{Or, more correctly, the reduced shear, $g \approx \gamma/(1-\kappa)$. For WL, $\kappa \ll 1$ and $g\approx\gamma$.} Convergence, a measure of the magnification of the image, can be reconstructed from the shear. In CMBWL both shear and convergence can be reconstructed from the temperature map. The analytic expressions given in this section are equally applicable to shear or convergence power spectra. 


Since their means vanish, it is convenient to quantify the fluctuations in both GWL and CMBWL with angular two-point functions, in particular auto- and cross-power spectra in harmonic space. Under the Limber approximation \citep{K92}, these take the form of integrals over the non-linear matter power spectrum, $P_{\delta\delta}(\ell/\chi(z),z)$, and a pair of appropriately chosen window functions. We are interested in the cross-correlation between GWL and CMBWL,
\begin{equation}
    \begin{split}
    C_{\rm GWL,CMBWL}(\ell) = \hspace*{5.0cm} \\
    \hspace*{-0.3cm} \int_{0}^{\chi_{\ast}} \frac{d\chi}{\chi(z)^{2}} W_{{\rm GWL}}\left[\chi(z)\right] W_{{\rm CMBWL}}[\chi(z)] P_{\delta\delta}\left(\frac{\ell}{\chi(z)},z\right),
    \end{split}
\label{eqn:theory_cl}
\end{equation}
where $\chi(z)$ is the comoving distance to redshift $z$, and $\chi_{\ast}$ is the distance to the horizon. Here $W_{\rm GWL}$ and $W_{\rm CMBWL}$ are the GWL and CMBWL window functions.

The GWL window function, also known as the lensing efficiency function or lensing kernel, takes the form
\begin{equation}
W_{{\rm GWL}}\left[\chi(z)\right] = \frac{3H_{0}^{2}\Omega_{\rm m}}{2c^{2}}\frac{\chi}{a(\chi)}\int_{\chi}^{\chi_{\ast}} d\chi' n(\chi')\frac{\chi'-\chi}{\chi'},
\label{eq:w_gwl}
\end{equation}
where $H_{0}$ is the Hubble parameter, $c$ the speed of light, $\Omega_{\rm m}$ the total matter density and $n(\chi')$ is the galaxy redshift distribution. We have assumed a flat universe, as we will throughout the paper.

The CMBWL window function takes a similar form but is somewhat simpler due to the single source plane,
\begin{equation}
W_{{\rm CMBWL}}\left[\chi(z)\right] = \frac{3H_{0}^{2}\Omega_{\rm m}}{2c^{2}}\frac{\chi}{a(\chi)}\frac{\chi_{*}-\chi}{\chi_{*}},
\label{eq:w_cmb}
\end{equation}
where $\chi_{*}$ is the comoving distance to the last scattering surface. Although the CMBWL weight function peaks at $z {\sim} 2$ \citep{LC06}, it is sensitive to the integrated gravitational potential between the source and the observer. The window functions for CMBWL and GWL corresponding to the DES sources used in this work are shown in \Cref{fig:kernels}. The DES source galaxies are sensitive to mass fluctuations at lower redshift than the peak of the CMB lensing kernel ($z {\sim} 2$) 
but there is sufficient overlap to expect a significant cross-correlation in the gravitational lensing signal of both sources.  

During the epoch of structure formation, galaxies experience tidal forces due to the gravitational potential of the surrounding mass distribution. The presence of such forces may cause the ellipticity and orientation of neighbouring galaxies to become aligned \citep{HRH2000,CKB01,HS04}. This effect, known as intrinsic alignment (IA), produces a correlation between the intrinsic shapes of galaxies and will be present as an extra term in our GWL$\times$CMBWL cross-correlation measurement \citep{TI14,HT14,CDM+15},
\begin{equation}
C^{\rm obs}_{{\rm GWL},{\rm CMBWL}}(\ell) = C_{{\rm GWL},{\rm CMBWL}}(\ell) + C_{{\rm IA},{\rm CMBWL}}(\ell).
\end{equation}

The IA$\times$CMBWL cross-correlation can be calculated, like the other power spectra, using \Cref{eqn:theory_cl} with the appropriate IA weight function instead of the GWL weight function. In this paper we assume IAs are described by the widely-used non-linear alignment (NLA) model \citep{BK07,HS10}, which means the weight function is given by
\begin{equation}
W_{\rm IA}\left[\chi(z)\right] = -C_1\rho_{\textrm{crit}}\frac{\om}{D\left[\chi(z)\right]}n\left[\chi(z)\right].
\label{eqn:IA_window}
\end{equation}
Here $\rho_{\rm crit}$ is the critical density at $z=0$ and $C_1 = 5 \times 10^{-14}h^{-2}M_{\odot}^{-1}{\rm Mpc}^3$, a normalisation constant based on the SuperCOSMOS measurement at low redshift \citep{brown_spercosmos_2002}. $D[\chi(z)]$ is the linear growth function, normalised to unity at $z=0$. Current measurements of IAs for different galaxy types over different redshifts still leave a significant uncertainty as to the expected level of the IA contribution for any given sample of source galaxies. The DES SV cosmology analysis \citep{deswl2ptcosmo2015} was consistent with an IA signal between zero and four times $C_1$ within 2$\sigma$ confidence limits. We treat the range of possible IA contributions in \Cref{sec:sys_IAs} and discuss the significance of IAs for our measurement and future GWL$\times$CMBWL analyses in \Cref{sec:discussion}. First though, we neglect the IA contribution in our main analysis, fitting for the pure GWL$\times$CMBWL signal only. Note that the negative sign in \Cref{eqn:IA_window} means that the IA contribution subtracts from the total observed GWL$\times$CMBWL cross-correlation. This is because the galaxy shape alignment sourced by IAs is of the opposite sense to that sourced by GWL. This means any measured cross-correlation amplitude will be lower than it would be if IAs were taken into account and that the IA contribution, if significant, would therefore {\it increase} the significance of the detection if it were included in the analysis.

\section{Data}\label{sec:data}

In this section we describe the data sets used in this paper. The DES SV data products are introduced in \Cref{sec:data_des}, including the photometric redshift (photo-$z$) and shear catalogues. Next, we describe the CMB data sets and their lensing $\kappa_{\rm CMB}$ maps, including those from SPT (\Cref{sec:data_cmb_spt}) and \emph{Planck} (\Cref{sec:data_cmb_planck}). Finally, in \Cref{sec:data_simulations}, we describe the GWL and CMBWL simulations and mock catalogues used for pipeline testing and covariance estimation.


\subsection{The Dark Energy Survey}\label{sec:data_des}

The Dark Energy Survey (DES) is an optical survey, currently in progress, which will cover 5,000 deg$^{2}$ in five filters ($grizY$) using DECam \citep{FDH+15} over five years, reaching a $10\sigma$ limiting magnitude of ${\sim}24.1$ in the \emph{i}-band \citep{sanchezea_desphotoz_2014}. The DES footprint was designed to overlap significantly with the region observed by the SPT (described in the next section), enabling many interesting cross-correlation measurements \citep{CarlstromSPT2011,giannantonia_ea_2015,saro2015}. In this paper we use the DES Science Verification (SV) data, which was taken during the period November 2012 -- February 2013, before the start of the main survey in late August 2013. Specifically, we use the 139 deg$^{2}$ contiguous area of the DES SV data that overlaps with the SPT East field; this is known as the SPT-E region \citep{jarvisetal2015} and is centred on RA $\sim$ 77.5 deg, DEC $\sim$ -51 deg. 
All the DES data products in this paper have been reduced from the raw survey data by the DES Data Management pipeline \citep{DESDM2012,Desai2012}.


\subsubsection{The DES SV photo-$z$ catalog}
\label{sec:data_des_photoz}

Large optical surveys like DES use photometry to estimate the redshift of source galaxies; this estimation technique is known as photo-$z$ \citep{HAC+10}. Fluxes in multiple broad filter bands are measured as a kind of very low resolution spectrum and a variety of methods are employed to estimate the corresponding true redshift, exploiting broad features of the spectral distribution rather than spectral lines. This is obviously less accurate than spectroscopic approaches, which produce high resolution spectra for each object, but the photo-$z$ approach is faster and cheaper for large imaging surveys. For our cross-correlation measurement it is essential that the overall redshift distribution, $n(z)$, is well characterised as this affects the theoretical cross-correlation power spectrum that we fit to our measured values.

The primary photo-$z$ catalogue used in this work is produced by the {\tt SkyNet2} neural network algorithm described in \cite{Graff2013,Bonnett2015,bonnettetal2015a}. Four different photo-$z$ methods were extensively tested and characterised in \cite{bonnettetal2015a} for the galaxy sample used in the DES SV shear catalogs. In addition to {\tt SkyNet2}, these were the {\tt ANNz2} \citep{SAL15}, {\tt BPZ} \citep{Benitez2000,Coe2006}, and {\tt TPZ} \citep{Carrasco2013, Carrasco2014} photo-$z$ estimation pipelines, selected because they performed well in the analysis of \cite{sanchezea_desphotoz_2014}. \cite{bonnettetal2015a} found that the catalogues agreed to better than $\Delta z < 0.05$ in the mean photo-$z$ of the distribution and we explore the effect of using these alternate photo-$z$ estimators in \Cref{sec:sys_photoz}.

\begin{figure}
\begin{center}
\hspace{-0.2cm}\includegraphics[width=0.48\textwidth]{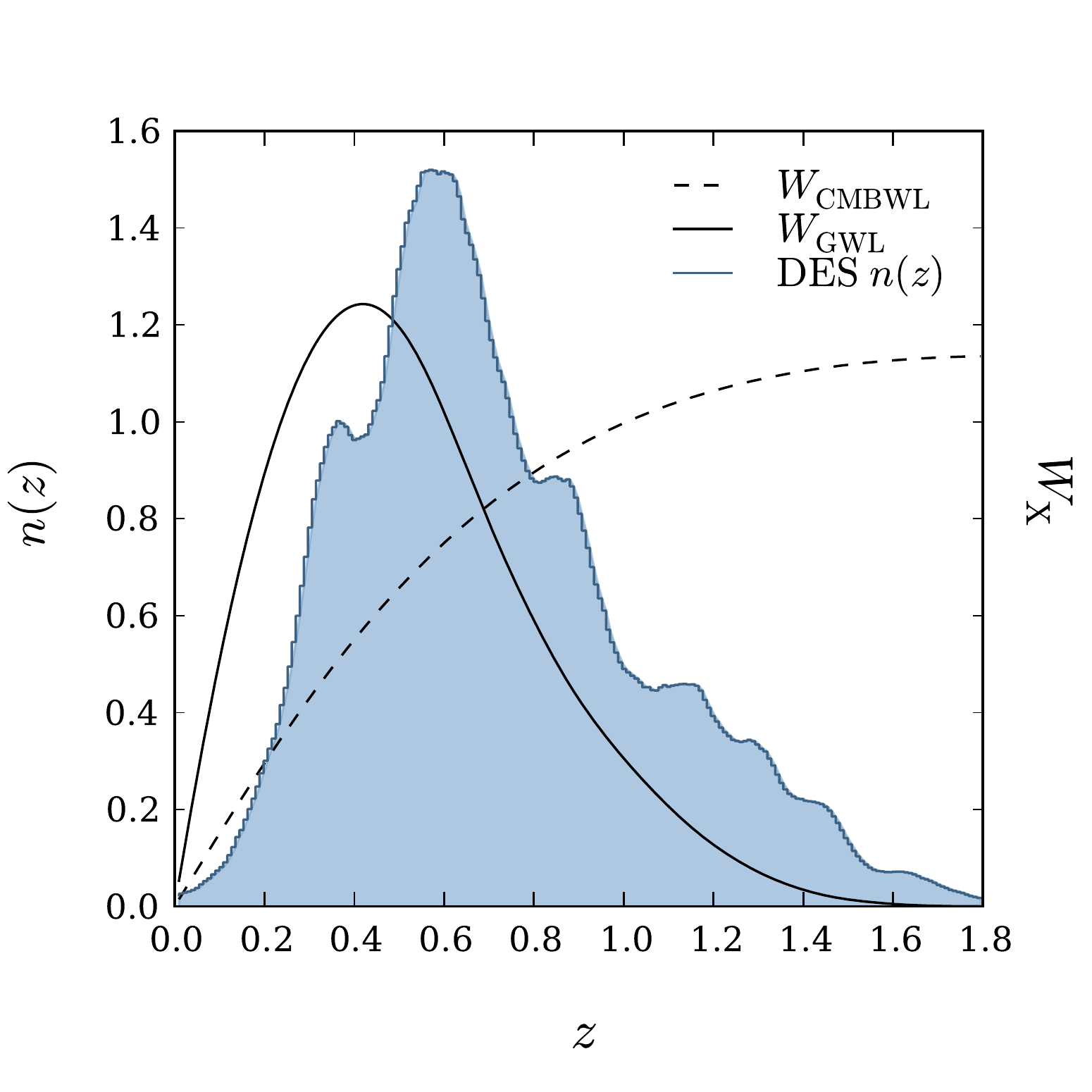}
\caption{The blue region shows the redshift distribution of the source galaxies from DES over our chosen redshift range, $0.3 < z < 1.3$, normalised such that the total area under the curve equals unity. The solid and dashed black lines show the lensing kernels for GWL and CMBWL respectively. Each weight function has been normalised for visual comparison with the $n(z)$. The DES sources are taken from the \ngm\ shape catalogue and we use the ${\tt SkyNet2}$ photo-$z$ estimates; see \Cref{sec:data_des_photoz,sec:data_des_shear} for more details.}
\label{fig:kernels}
\end{center}
\end{figure}

{\tt SkyNet2} produces a full probability density function, $p(z)$, for each galaxy. We use the mean of the $p(z)$ for each galaxy, $z_{\rm mean}$, as its point-estimate redshift and select/bin the galaxies according to $z_{\rm mean}$. We use this point estimate to select galaxies in the redshift range $0.3<z<1.3$, as this is the range over which the {\tt SkyNet2} algorithm gives reliable photo-z estimates according to \citet{bonnettetal2015a}. Other point estimates were tested but the mean produced the best results for our application as it reduces the impact of unphysical features in individual galaxy $p(z)$. However, for the calculation of the theoretical prediction of our cross-correlation measurements, we use the full stacked $p(z)$ from each galaxy to estimate the total redshift distribution for our forecast.  
\Cref{fig:kernels} shows the resulting redshift distribution of the DES sources used in this cross-correlation as well as the corresponding GWL lensing kernel, \Cref{eq:w_gwl}, and the CMBWL kernel, \Cref{eq:w_cmb}, for comparison.

\subsubsection{The DES SV shear catalog}
\label{sec:data_des_shear}
\vspace{-0.2cm}The DES shear catalogs, described in detail by \citet{jarvisetal2015}, are based on two independent galaxy shape measurement algorithms: \ngm\ \citep{S14} and \ims\ \citep{zuntz2013}. Note that although the coadd catalogues were used in the initial processing of the shear measurement pipeline, the galaxy shape measurements were carried out at the single-exposure level. The shape (or shear) of each galaxy is estimated by jointly fitting a galaxy model to multiple single-exposure images of that same galaxy \citep{jarvisetal2015}, thus reducing the impact of instrumental artefacts and PSF variation between exposures. 

The analysis of \citet{jarvisetal2015} showed that both catalogues passed all requirements on contamination by systematic effects and that the two catalogues were consistent with each other under a range of statistics. \citet{beckeretal2015} and \citet{deswl2ptcosmo2015} demonstrated that the two catalogues remain consistent at the level of two-point statistics and inferred cosmological constraints. We use the \ngm\ catalogue for our main analysis because it has a larger effective source galaxy number density, 5.7 arcmin$^{-2}$ as opposed to 3.7 arcmin$^{-2}$ for \ims\ (see \citet{jarvisetal2015} for details of the weighting used to calculate this effective number density). 
This choice is consistent with the DES SV GWL two-point analysis \citep{beckeretal2015} and the DES SV GWL cosmology analysis \citep{deswl2ptcosmo2015}. We repeat our analysis using the \ims\ catalogue as a consistency check; these results can be found in \Cref{sec:sys_im3shape}. 
Both catalogues have demonstrated that they can provide shear measurements with systematic uncertainties (whether from astrophysical, observational, or measurement effects) subdominant to the statistical uncertainty in the SV data for different cosmological probes including two-point statistics in real and harmonic space, galaxy-galaxy lensing and mass-mapping \citep{jarvisetal2015,changetal2015,VCJ+15,beckeretal2015,clampittetal2015}.

We produce maps of our shear catalogues, described in more detail in \Cref{sec:results_GKharmonic_implementation}, using the \texttt{HEALPix} pixelisation scheme at $N_{\text{side}}=2048$ \citep{gorski2005}. This corresponds to a pixel area of $2.95$ arcmin$^{2}$ or a pixel scale of ${\sim} 600$\ kpc at $z=0.44$, our redshift of maximal sensitivity.

Shape estimates from both pipelines were `blinded' during our analysis to avoid experimenter bias \citep{KR05}. 
This meant that a constant scaling factor (between 0.9 and 1) was applied to all ellipticities. This would slightly alter the amplitude of the cross-correlation, preventing over-fitting to results from other papers or to any given cosmology. 
Our analysis procedure was finalised and fixed before de-blinding.

\subsection{CMBWL Maps}
\label{sec:data_cmb}

\subsubsection{SPT Lensing Maps}
\label{sec:data_cmb_spt}

The CMB $\kappa$ maps are based on temperature measurements made with the SPT \citep{CarlstromSPT2011}, which is a 10 metre diameter telescope located at the National Science Foundation Amundsen-Scott South Pole Station in Antarctica. During 2008-2011, this telescope was used to conduct a tri-band (90, 150, 220 GHz) wide-field survey covering ${\sim}$ 2540 ${\rm deg}^{2}$ \citep{SRH+13}. The survey area is composed of 19 subfields, all of which were scanned in a similar fashion, reaching minimum depths (maximum noise levels) of 40 $\mu {\rm K}$-arcmin (90 GHz), 18 $\mu {\rm K}$-arcmin (150 GHz) and 70 $\mu {\rm K}$-arcmin (220 GHz) with roughly arcminute resolution. 

The $\kappa_{\rm CMB}$ maps are produced by applying quadratic estimators \citep{OH03} on the filtered 150 GHz temperature map in the range $\ell^{\rm temperature}<4000$ on a $25^{\circ}\times25^{\circ}$ region extracted from the full survey area, centred on the DES SPT-E field. Modes with $\ell_{x}^{\rm temperature} < 500$ are also removed to minimise extra noise along the scanning direction \citep{vanengelen12}. Positive sources brighter than $15\sigma$ (corresponding to approximately 10 mJy) and clusters detected above 6$\sigma$ were masked with a $16' \times 16'$ aperture, and the masked regions were filled using Wiener filter interpolation. The maps are produced on a \texttt{HEALPix} grid of $N_{\text{side}}=2048$.

The signal-to-noise in our cross-correlation drops significantly at high $\ell$, and the contamination to the $\kappa$ reconstruction from emissive sources and galaxy clusters that are not masked becomes increasingly important at high $\ell$. For these reasons, we impose a conservative cut on the $\kappa$ map of $\ell_\mathrm{max}= 1600.$ This cut does not lead to a significant loss of signal. We have tested this choice of $\ell_\mathrm{max}$ and found our result robust to values between $\ell=1200$ and $\ell=2000$.  All of these issues are even more relevant to the \emph{Planck} CMBWL data (which has higher noise per mode than SPT, particularly at high $\ell$, as well as potentially higher contamination from point sources and clusters because of the larger \emph{Planck} beams), and we apply the $\ell_{\rm max}$ cut to the \emph{Planck} data (described in the next section) as well.

\subsubsection{Planck}\label{sec:data_cmb_planck}
We use the \emph{Planck} lensing maps from the second data release, which were made public in 2015\footnote{http://pla.esac.esa.int/pla/}. These $\kappa$ maps are produced by using filtered temperature and polarisation measurements from the \emph{Planck} satellite \citep{planck15xv}. The temperature and polarisation maps are both constructed by taking linear combinations of multi-frequency data (30\ -\ 857\ GHz for temperature and 30\ -\ 353\ GHz for polarisation) with scale-dependent coefficients using the {\tt SMICA} method to produce foreground cleaned minimum variance maps \citep{planck15ix}. \emph{Planck} HFI beams range from 4 (857\ GHz) to 10 (100\ GHz) arcminutes in resolution, compared to SPT's 1 arcminute. \emph{Planck} covers the full sky, while SPT is focused on a smaller patch.

Similar to the SPT $\kappa$ map, the CMBWL potential is produced using the quadratic estimators from \citet{OH03}. The main difference here is the availability of E and B-mode polarisation, which allows for additional estimators ($\phi^{TE}$, $\phi^{EE}$, $\phi^{EB}$, $\phi^{TB}$) in addition to $\phi^{TT}$. These estimators are combined to form a minimum-variance estimate of the lensing potential $\phi$, which is provided in the form of spherical harmonic coefficients of the CMB lensing convergence $\kappa$, filtered to $8\le \ell < 2048$, along with the analysis mask. The map is in \texttt{HEALPix} format with resolution $N_{\text{side}}=2048$.

\subsection{Simulations}\label{sec:data_simulations}

We test our estimators using simulated data sets, constructed specifically to mimic the noise and other statistical properties of each of our observables: GWL shape catalogues for DES and CMBWL convergence maps for SPT and \emph{Planck}.

For DES, we use two sets of mock catalogues in addition to the data itself. The first are based on N-body simulations and are the same set of simulated galaxy catalogues described in \citet{beckeretal2015}, consisting of 126 realisations of the SV SPT-E patch with lensing fields calculated by ray-tracing. 
We sub-sample the galaxies in the simulations to match the galaxy number density, redshift distribution and noise properties of the data, drawing a corresponding shape value from the data and adding it to the cosmological shear signal from the simulation to ensure a realistic shape noise distribution.

When estimating the noise properties of our cross-correlations we also employ a separate set of 100 DES mock catalogues produced, not from N-body simulations, but simply by applying a random rotation to the orientation of each source galaxy in our DES shape catalogue. This retains the spatial and redshift distribution of galaxies as well as the overall intrinsic ellipticity distribution across the sample, while destroying any cosmological information. Therefore these randomised catalogues act as noise-only realisations.


The SPT collaboration has produced 100 lensed and 100 unlensed simulated sky realisations. Instrumental noise is added to these sky realisations and the result is converted into simulated time streams, then processed in the same way as the real data, including the effects of masking and filtering. The 100 output reconstructions were then cross-correlated with the input lensing potential, $\phi$, to obtain the lens transfer function. This transfer function was then applied to $\kappa$ reconstructions from the unlensed CMB realisations, providing 100 realistic realisations of noise in the $\kappa$ map. Noise in the CMB convergence maps comes primarily from fluctuations in the CMB temperature field itself, rather than instrumental noise. More discussion of these noise simulations can be found in \citet{giannantonia_ea_2015}.
%
%


\emph{Planck} has produced 100 simulated lensing maps using their Full Focal Plane 8 (FFP8) Monte Carlo simulations \citep{planck_fullfocalplanesims_2015}. 
The noise level in these simulations has been tuned by the \emph{Planck} collaboration to match the amplitude of the \emph{Planck} CMB power spectra. 

\section{Methods}\label{sec:results_GKharmonic_implementation}

In this section we describe our main analysis pipeline - how we make our cross-correlation measurement and calculate the associated error. This main analysis is conducted using the \texttt{PolSpice} code \citep{SPP+01,chon2004} to estimate projected angular power spectra in harmonic space using the \ngm\ shape catalogues and the CMB convergence maps from SPT and \emph{Planck}. The results themselves are presented in \Cref{sec:results}. 



We produce shear maps by averaging shear estimates for individual galaxies from the DES shape catalogues into \texttt{HEALPix} pixels at a resolution of $N_{\text{side}}=2048$, applying the standard cuts, weighting and bias corrections as described in \citet{jarvisetal2015}. The two shear components, $\gamma_1$ and $\gamma_2$, are treated separately to produce two maps. We then combine $\gamma_1$ and $\gamma_2$ maps from DES with the $\kappa_{\rm CMB}$ maps from SPT/\emph{Planck} into a $\{\kappa_{\rm CMB},\gamma_{1},\gamma_{2}\}$ triplet. We use the \texttt{PolSpice} \citep{SPP+01,chon2004} code in polarisation mode to estimate projected angular power spectra for the auto- and cross-correlations of $\kappa_{\rm CMB}$ and DES $\gamma$. Our input $\{\kappa_{\rm CMB},\gamma_{1},\gamma_{2}\}$ are analogous to the CMB $\{T,Q,U\}$ triplet. We have tested this process 
on the DES, SPT and \emph{Planck} simulations described in \Cref{sec:data_simulations} to confirm it is capable of returning an unbiased estimate in the presence of the relatively restrictive DES SV survey mask. 

The shear maps can be decomposed into spin $\pm2$ spherical harmonics
\begin{equation}
\frac{1}{2}(\gamma_{1}(\hat n)+i\gamma_{2}(\hat n))=\sum_{\ell m } p_{\pm 2, \ell m}\ _{\pm 2}Y_{\ell m},
\end{equation}
where $\ _{\pm 2}Y_{\ell m}$ and $p_{\pm 2, \ell m}$ are the spin $\pm2$ spherical harmonics and their coefficients. Of particular interest to us are the linear combinations of these spherical harmonics which produce curl-free E-mode and divergence-free B-mode components of the shear field \citep{bartelmann10,SvW+02,CN+02}:
\begin{align}
\gamma_{{\rm E},\ell m} =&\ - (p_{+2, \ell m} + p_{-2, \ell m}),\\
\gamma_{{\rm B},\ell m } =&\ -i(p_{+2, \ell m} - p_{-2, \ell m}).
\end{align}
To first order, WL only generates E-modes because the lensing potential is a real scalar \citep{SvW+02,CN+02}. We therefore aim to measure the $\kappa_{\rm CMB}\gamma_{\rm E}$ cross-correlation. Any measurable $\kappa_{\rm CMB}\gamma_{\rm B}$ correlation is unphysical and  evidence of some untreated systematic effect in the DES data which is correlated with CMBWL (see \Cref{sec:sys_Bmodes} below for $\kappa_{\rm CMB}\gamma_{\rm B}$ analysis). The decomposition is handled by \texttt{PolSpice}, which is designed to make the same split into curl- and divergence-free components in the context of CMB polarisation studies \citep{MCD+02}. We have used our simulation catalogues to confirm that \texttt{PolSpice} can recover an unbiased estimate of the cosmic shear power spectrum via this method.

We apply simple binary masks to our data. For DES, \texttt{HEALPix} pixels are set to zero if they do not contain any DES SV source galaxies after all our quality cuts have been applied; the remaining region covers 139 deg$^2$. For SPT we have data over a wider contiguous patch of 600 deg$^2$, overlapping the DES SV SPT-E region. The mask is one inside this region and zero outside. For \emph{Planck} we use the publicly available mask designed for power spectrum estimation, where point sources and galactic emission are masked out, removing roughly one third of the full sky. 

We take a hybrid approach to estimating the noise in our cross-correlation, describing the noise with an analytic expression that includes estimates of noise in the constituent GWL and CMBWL parts acquired from simulations or randomised realisations of the data. Alternate noise estimators and the considerations governing our choice are described in \Cref{sec:sys_covmats}. The analytic form of the noise is given by 
\begin{equation}
    \begin{split}
    \hspace{-0.2cm}\sigma^{2}_{\kappa_{\rm CMB}\gamma_{\rm E}}(\ell) = \left( \frac{1}{f_{\rm sky}(2\ell + 1)\Delta\ell} \left[ C_{\kappa_{\rm CMB}\gamma_{\rm E}}(\ell)C_{\kappa_{\rm CMB}\gamma_{\rm E}}(\ell) + \hspace{0.75cm} \right. \right. \\
    \left. \left. (C_{\kappa_{\rm CMB}\kappa_{\rm CMB}}(\ell)+N_{\kappa_{\rm CMB}\kappa_{\rm CMB}}(\ell))(C_{\gamma_{\rm E}\gamma_{\rm E}}(\ell)+N_{\gamma_{\rm E}\gamma_{\rm E}}(\ell)) \right] \vphantom{\frac{1}{f_{\rm sky}(2\ell + 1)\Delta\ell} }\right),\hspace{0.5cm}
    \end{split}
\label{eqn:error}
\end{equation}
where $C_{\gamma_{\rm E}\gamma_{\rm E}}$, $C_{\kappa_{\rm CMB}\kappa_{\rm CMB}}$ and $C_{\kappa_{\rm CMB}\gamma_{\rm E}}$ are the theory power spectra of the WL auto-correlation, the CMB lensing auto-correlation and the WL/CMB cross-correlation respectively, assuming the \emph{Planck} 2015 best-fit cosmology, and $f_{\rm sky}$ is the fraction of sky covered by the survey. These are calculated using CAMB sources \citep{CL11}. 

The auto-correlation contributions to the noise estimate include measurement noise terms, $N_{\gamma_{\rm E}\gamma_{\rm E}}(\ell)$ and $N_{\kappa_{\rm CMB}\kappa_{\rm CMB}}(\ell)$. The noise term for DES is the mean of the auto-correlations of the maps produced from the 100 randomised realisations of our shape catalogue. The CMBWL noise term is calculated as the mean of the auto-correlations of the 100 noise realisations provided by both the SPT and \emph{Planck} collaborations. Each of these sets of noise simulations are described in \Cref{sec:data_simulations} above. Note that we have estimated the uncertainty using the mean noise level from \emph{Planck}, whereas in reality  the SV patch covers a region where Planck noise is lower than average, hence we are slightly over-estimating the noise in the DES$\times$\emph{Planck} case. 

We calculate $f_{\rm sky}$ using our most restrictive mask, in this case that of the DES shape catalogues. As mentioned above, the DES SV SPT-E patch which we are using covers 139\ deg$^2$ of sky, though the exact sky fraction varies slightly depending on the shape catalogue and redshift range considered. We take this variation into account in our calculations. In the case of DES$\times$\emph{Planck} there are a small number of additional regions inside the DES SV mask but excluded from the \emph{Planck} mask. We take the product of the two masks when estimating $f_{\rm sky}$, reducing the sky fraction by 1.5\% compared to DES$\times$SPT. 


We also tested a number of alternative techniques to estimate the cross-correlation signal, covariance and noise as consistency checks on our main analysis. These are detailed in \Cref{sec:systematics} below.

\section{Results}\label{sec:results}
\begin{center}
\begin{figure}
\begin{tabular}{cc}
\includegraphics[width=0.48\textwidth]{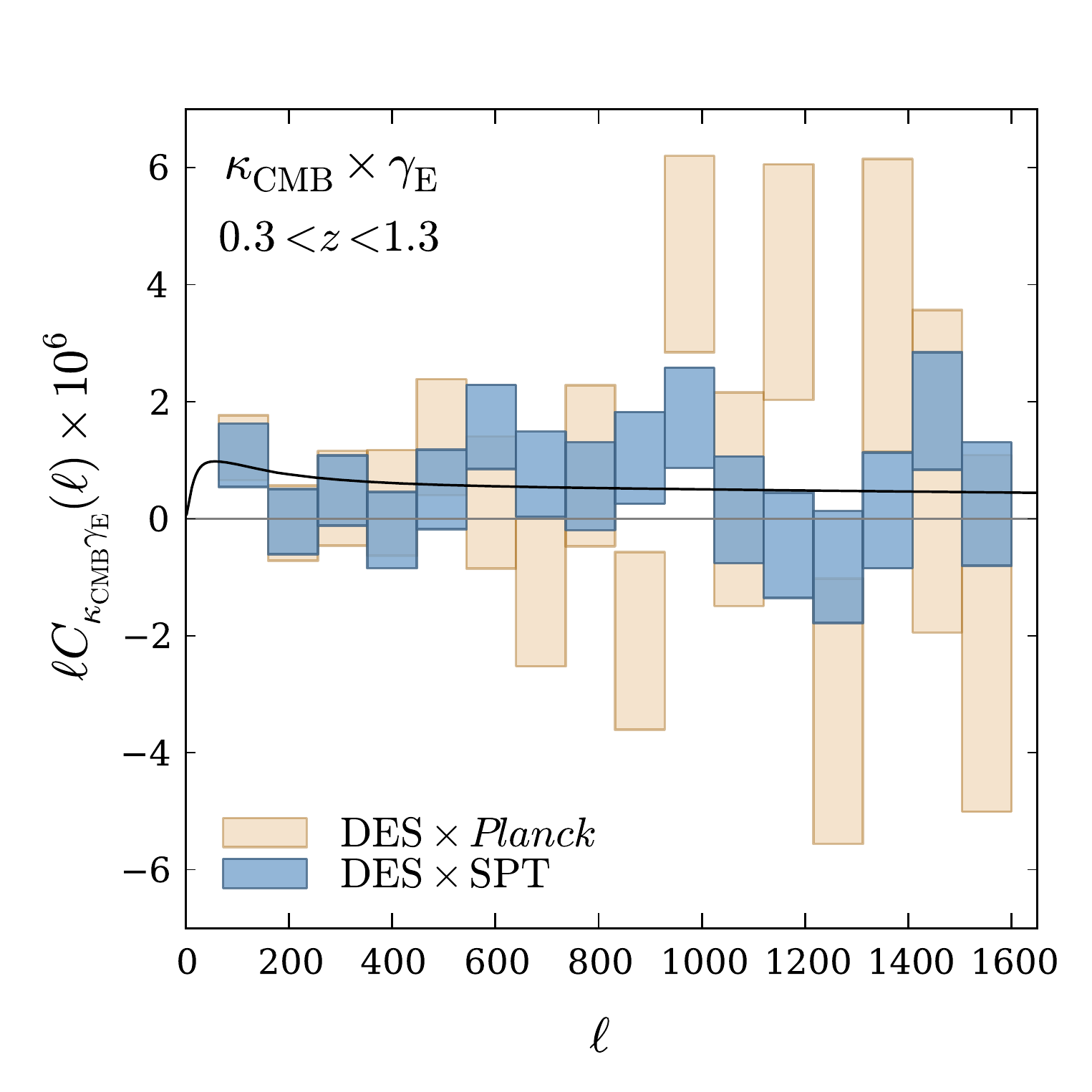}
\end{tabular}
\caption{$\kkE$ correlation measured in harmonic space with \texttt{PolSpice}. Projected angular power spectra, $C(\ell)$, are shown for DES $\times$ SPT (blue boxes) and DES $\times$ \emph{Planck} (orange boxes). Sources come from the \ngm\ shape catalogue and span the redshift range $0.3<z<1.3$, selected using the ${\tt SkyNet2}$ photo-$z$ catalogue. The height of the bars represents 68\% error limits. The theoretical prediction for the cross-correlation, with amplitude $A=1$, is also shown (black solid line).} 
\label{fig:Emode_results_GK_harmonic}
\end{figure}
\end{center}
\vspace{-0.5cm}\Cref{fig:Emode_results_GK_harmonic} shows our results using \texttt{PolSpice} to correlate $\kappa_{\rm CMB}$ with $\gamma_{\rm E}$ in harmonic space. The measurement is averaged into 16 linearly-spaced bins over the multipole range $64 < \ell < 1600$. We then use our forecast cross-correlation power spectrum to fit a single free parameter, the cross-correlation amplitude $A$, by a simple $\chi^2$ minimisation:
\begin{equation}
\chi^{2} = \sum_\ell \frac{\left(C^{\rm obs}_{\kkE}(\ell) - A \times C^{\rm theory}_{\kkE}(\ell)\right)^{2}}{\sigma^{2}_{\kappa_{\rm CMB}\gamma_{\rm E}}},
\end{equation}
where the error, $\sigma_{\kappa_{\rm CMB}\gamma_{E}}$, is calculated according to \Cref{eqn:error}. 
The fits to the cross-correlation amplitude are detailed in \Cref{table:results}. If our measurement were consistent with the expectation from theory, assuming the \emph{Planck} cosmology and that we have correctly modelled the DES galaxy redshift distribution, then we would expect a result consistent with $A=1$.


\begin{table}
\centering
\begin{tabular}{l|cc}
\hline
Redshift Range & \multicolumn{2}{c}{$0.3 < z < 1.3$}  \\
\hline
{\bf $\kkE$} & $A$ & $\chi^{2}/dof$ \\
\hline
\ngm\ $\times$ SPT              & $0.88 \pm 0.30$ & 0.93\\
\ngm\ $\times$ \emph{Planck}    & $0.86 \pm 0.39$ & 1.52\\
\hline
\end{tabular}
\caption{Summary of constraints on the cross-correlation, $\kkE$, showing best-fit cross-correlation amplitude, $A$, with $1 \sigma$ errors and minimum $\chi^{2}/dof$ (where $dof = 15$). Results are shown for cross-correlations between DES GWL from the \ngm\ catalogue and CMBWL from both SPT and \emph{Planck}.}
\label{table:results}
\end{table}



Our measurement shows E-mode cross-correlations with best-fit amplitudes of $A=0.88\pm0.30$ for DES$\times$SPT and $A=0.86\pm0.39$ for DES$\times$\emph{Planck}, giving a significance of $2.9\sigma$ and $2.2\sigma$ respectively. We estimate the goodness-of-fit by calculating $\chi^{2}$ per degree of freedom (15, the number of $\ell$ bins minus one), finding good fits in both cases, with $\chi^{2}/dof=0.93$ for DES$\times$SPT and $\chi^{2}/dof=1.52$ for DES$\times$\emph{Planck}. The measurements with SPT and \emph{Planck} are consistent with each other and with the theoretical expectation. The DES$\times$\emph{Planck} cross-correlation has a relatively high $\chi^{2}/dof$, with the probability to exceed such $\chi^2$ being $\sim 10\%$; a similar result was found in \citet{giannantonia_ea_2015} for DES SV LSS$\times$\emph{Planck} CMBWL. These measurements of $A$ fix all other cosmological parameters at the \emph{Planck} 2015 best-fit cosmology and ignore IAs.

We consider in detail the impact of a variety of systematic effects in \Cref{sec:systematics}. Some effects, including uncertainty in shear measurement bias and estimation of photo-z could change the measured amplitude of the cross-correlation. The result quoted here should be considered the `bare' constraint on $A$, when we assume our best estimates for shear measurement bias, photo-z. This is a reasonable approach because any deviation in these quantities will scale both the best-fit amplitude and the error bars, leaving the significance of detection unaffected. 

\section{Consistency and Systematics Tests}\label{sec:systematics}
In this section we summarise a number of checks carried out to ensure that our analysis is accurate and robust to observational and astrophysical systematic effects.
A substantial amount of work has been done quantifying the contribution of systematics to our data sets \citep{jarvisetal2015,bonnettetal2015a,beckeretal2015,VCJ+15,leistedtea2015}.
In particular, \citet{giannantonia_ea_2015} dealt with a number of systematics that could potentially manifest as spurious signal in the cross-correlation of DES SV galaxy number density with CMBWL; all were found to be of negligible importance. In this paper we will concentrate on those of particular relevance to the GWL$\times$CMBWL cross-correlation. 


\subsection{Shape Measurement Pipelines}
\label{sec:sys_im3shape}

As mentioned in \Cref{sec:data_des_shear}, the DES collaboration has produced two independent shape measurement catalogues: \ngm\ \citep{S14}, 
which we use for our main analysis, and \ims\ \citep{zuntz2013}.

We have repeated our cross-correlation measurement using the \ims\ shape catalogue. The results are in good agreement with those from \ngm\ and our forecasts but the errors are larger due to the lower effective source number density in \ims\ (3.7/arcmin$^{2}$ compared to 5.7/arcmin$^{2}$ for \ngm). 
With \ims, we measure a cross-correlation amplitude of $A = 0.76 \pm 0.38$ for DES$\times$SPT and $A = 0.76 \pm 0.53$ for DES$\times$\emph{Planck}. 
Like our main results, the \ims\ measurements are consistent with the expected signal within $1\sigma$ errors. 

\subsection{Alternate Estimators}

Besides the aforementioned {\tt PolSpice} pipeline, we tested an additional flat-sky implementation of the same calculation, also known as the Kaiser-Squires method \citep[KS,][]{ks93}. The KS method was used in both \citet{HLD+13} and \citet{LH15} and shown to perform well. With an eye on the larger sky coverages in future data sets, we have performed our main calculations based on a curved sky analysis but we also checked whether our results are consistent with the those from the flat-sky KS analysis used in previous literature. 

In the flat-sky KS approach $\gamma_{1}^{\rm cat}$ and $\gamma_{2}^{\rm cat}$ for each galaxy in the shear catalogue are projected onto a zenith equal area projection coordinate grid, then averaged over all the galaxies that fall on the same square grid. Since the shear measurements are made with respect to spherical coordinates, we apply the rotation 
\begin{align}
\gamma_{1}&={\rm cos}(2\varphi)\gamma_{1}^{\rm cat}-{\rm sin}(2\varphi)\gamma_{2}^{\rm cat}\\
\gamma_{2}&={\rm sin}(2\varphi)\gamma_{1}^{\rm cat}+{\rm cos}(2\varphi)\gamma_{2}^{\rm cat},
\end{align}  
where $\varphi$ is the \emph{local} angle between equal right ascension to the $y$-axis of the image, such that the shear measurements are described with respect to the flat-sky $xy$-coordinates.

 The $\gamma$ maps are converted into $\gamma_{\rm E}$ by Fourier transforming $\gamma_{1},\gamma_{2}$ using 
\begin{equation}
\gamma_{\rm E}(\boldsymbol{\ell_{x}},\boldsymbol{\ell_{y}})=\gamma_{1}(\boldsymbol{\ell_{x}},\boldsymbol{\ell_{y}})\frac{\ell_{x}^2-\ell_{y}^2}{\ell_{x}^2+\ell_{y}^2}+\gamma_{2}(\boldsymbol{\ell_{x}},\boldsymbol{\ell_{y}})\frac{2 \ell_{x} \ell_{y}}{\ell_{x}^2+\ell_{y}^2},
\end{equation}
and are cross-correlated with CMB $\kappa$ in Fourier space,
\begin{equation}
{C}_{\kappa_{\rm CMB}\gamma_{\rm E}}(\ell)=\left\langle \kappa_{\rm CMB}(\boldsymbol{\ell_{x}},\boldsymbol{\ell_{y}}) (\gamma_{\rm E}(\boldsymbol{\ell_{x}},\boldsymbol{\ell_{y}}))^{*}\right\rangle_{\ell},
\end{equation}
where $\ell^{2}=\ell_{x}^{2}+\ell_{y}^{2}$. In this process, we also apply a mask due to the finite survey area, which inevitably induces mode-coupling between the bins. To account for the mask, we follow \citet{hivon02} to decouple the effect of the mask from the computed spectra. Using this method, we obtain amplitudes of $A=0.92 \pm 0.30$ for DES$\times$SPT and $A=0.91 \pm 0.39 $ for DES$\times$\emph{Planck}, with a $\chi^{2}/dof$ of 1.18 and 1.17 respectively, which is consistent with the {\tt PolSpice} pipeline.
 


We also tested an alternate method which uses the pseudo-Cl estimation technique for CMB polarization developed in \cite{kogut03}. Spectra are computed from pseudo-multipole coefficients and related to the true spectra through a coupling matrix, which we bin before inversion \citep[see also][]{efstathiou06}. Using this approach, we obtain amplitudes of $A=0.82\pm0.32$ for DES$\times$SPT and $A=0.91\pm0.39$ for DES$\times$\emph{Planck}, which are in good agreement with the other two methods.


\subsection{Alternate covariance estimates}
\label{sec:sys_covmats}

We have checked the hybrid noise estimates used in our main analysis by also estimating the full covariance of our cross-correlation using both our N-body simulations and jack-knife resampling. 
Both estimate the covariance according to
\begin{align}
\begin{split}
\rm{Cov}(C_{\kappa_{\rm CMB}\gamma_{\rm E}}(\ell_{i}),& C_{\kappa_{\rm CMB}\gamma_{\rm E}}(\ell_{j})) = \\ 
\hspace*{-1.3cm} f(N) \sum_{v=1}^{N}&(C^{v}_{\kappa_{\rm CMB}\gamma_{\rm E}}(\ell_{i})-\bar{C}_{\kappa_{\rm CMB}\gamma_{\rm E}}(\ell_{i}))\\\times &(C^{v}_{\kappa_{\rm CMB}\gamma_{\rm E}}(\ell_{j})-\bar{C}_{\kappa_{\rm CMB}\gamma_{\rm E}}(\ell_{j})),
\end{split}
\label{eqn:covmat}
\end{align}
where $f(N)=1/(N-1)$ for the N-body case and $f(N)=(N-1)/N$ for the jack-knife and $v$ counts over the $N$ separate cross-correlation realisations, ${C}^{v}_{\kappa_{\rm CMB}\gamma_{\rm E}}(\ell)$. For the N-body method $N=100$, the number of DES/SPT simulations used, and for the jack-knife $N=40$, the number of equal area regions (3.5\ deg$^2$) into which the data is split. $\bar{C}_{\kappa_{\rm CMB}\gamma_{\rm E}}(\ell)$ is the mean of the $N$ cross-correlations in each case. For more details of both approaches to covariance estimation see the excellent review by \citet{NBG+09}.

All three approaches give consistent estimates of the error, with both the jack-knife and N-body errors agreeing with our hybrid estimate to within 10\% at all $\ell$. There is little off-diagonal power in the resulting covariance matrices using either method. When inverting the covariance matrices we apply the $\beta$ correction of \citet{HSS07} to account for the effect of having a finite number of realisations. We are confident that our hybrid approach remains the most accurate available noise estimate for this cross-correlation measurement. The simulations used are not correlated between the GWL and CMBWL, therefore the N-body method, which would otherwise be preferred, misses power due to both correlated non-linear structure growth and correlated cosmological signal. In addition the assumed cosmology in the simulations produces a lower amplitude signal than the \emph{Planck} best-fit assumed throughout the rest of this work. The jack-knife approach is useful as a consistency check but is known to over-estimate errors on small scales \citep{NBG+09}. In future work, with the full DES survey data set, the production of correlated N-body simulations of GWL and CMBWL should be a main priority so that we can use the full covariance estimate and capture off-diagonal power from mode-mixing due to correlated structure formation and mask effects.

\subsection{B-modes}\label{sec:sys_Bmodes}

In \Cref{sec:results_GKharmonic_implementation} we describe how the GWL shear field, $\gamma$, can be decomposed into curl-free (E-mode) and divergence-free (B-mode) contributions. 
WL will only produce E-mode signal to first order so, in cosmic shear measurements, detection of a significant B-mode auto-correlation is a diagnostic for systematics in the observations or the measurement, for example a poorly reconstructed PSF. While we do not necessarily expect these processes to generate B-modes that are positively correlated with CMBWL, it is possible that our estimator could introduce spurious power to our measurement (which would appear equally in the E/B-mode) or allow power to leak from the $\kkE$ to the $\kkB$ signal. Testing for significant $\kkB$ cross-correlation is therefore still a useful check on the efficacy of our estimators. Note that although the CMB lensing signature may contain some B-mode fluctuations, the magnitude is substantially less than the gravitationally-induced lensing signal \citep{planck13xvii} and we ignore this effect in our current study.  

\begin{figure}
\includegraphics[width=0.48\textwidth]{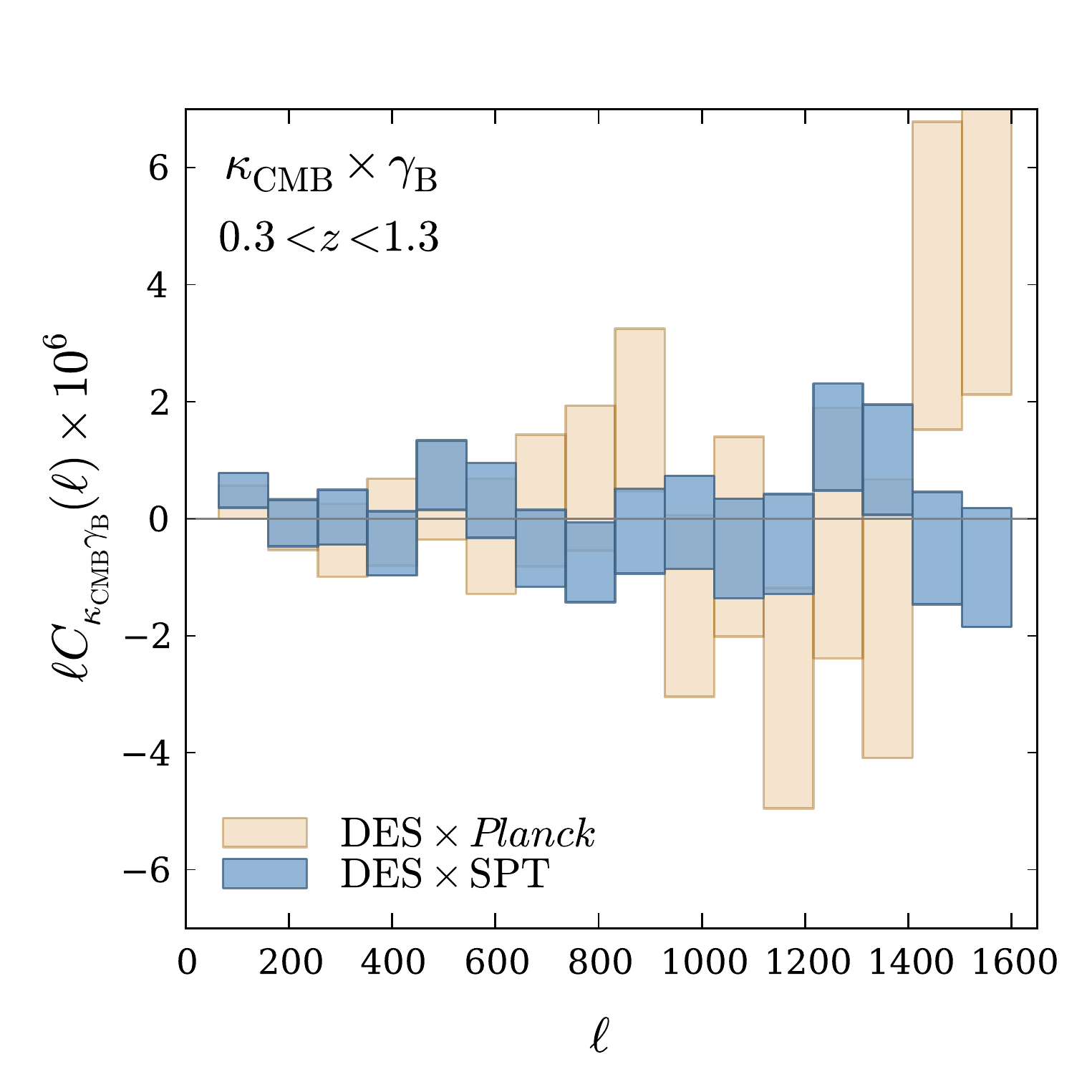}
\caption{$\kkB$ correlation measured in harmonic space with \texttt{PolSpice} for DES $\times$ SPT (blue boxes) and DES $\times$ \emph{Planck} (orange boxes). Sources come from the \ngm\ shape catalogue and span the redshift range $0.3<z<1.3$, selected using the ${\tt SkyNet2}$ photo-$z$ catalogue. The height of the bars represents 68\% error limits.}
\label{fig:Bmodes}
\end{figure}

In \Cref{sec:results} we presented the cross-correlation of our GWL E-mode signal with lensing from the CMB, $\kkE$. In \Cref{fig:Bmodes} and \Cref{table:results_Bmodes} we show the equivalent $\kkB$ cross-correlation. 
We estimate the significance of the $\kkB$ signal by fitting the expected E-mode signal with, as before, a varying constant cross-correlation coefficient, $A_{\rm B}$. We estimate the best-fit value of that constant by minimising $\chi^{2}$ in the same way as with the theoretical forecasts in our main $\kkE$ measurement. A significant signal of this kind would indicate leakage of power from E- to B-modes. Our best-fit B-mode cross-correlation amplitudes are consistent with zero for both DES$\times$SPT, $A_{\rm B} = 0.18 \pm 0.21$, $\chi_{\rm min}^{2}/dof=0.79$, and DES$\times$\emph{Planck}, $A_{\rm B} = 0.17 \pm 0.25$, $\chi_{\rm min}^{2}/dof=0.92$. We also checked the $\chi^{2}/dof$ for zero cross-correlation, $A_{\rm B}=0$, finding this to be a good fit to the data in both the DES$\times$SPT, $\chi_{A_{\rm B}=0}^{2}/dof=0.83$, and DES$\times$\emph{Planck}, $\chi_{A_{\rm B}=0}^{2}/dof=0.95$, cases. This gives us confidence that our estimator is not suffering from either spurious power or E to B leakage at a level that could bias our measurement significantly, given the size of our errors.

\begin{table}
\centering
\begin{tabular}{l|ccc}
\hline
Redshift Range & \multicolumn{3}{c}{$0.3 < z < 1.3$}  \\
\hline
{\bf $\kkB$} & $A_{\rm B}$ & $\chi_{\rm min}^{2}/dof$ & $\chi^{2}(A_{\rm B}=0)/dof$ \\
\hline
\ngm\ $\times$ SPT              & $0.18 \pm 0.21$ & 0.79 & 0.83\\
\ngm\ $\times$ \emph{Planck}    & $0.17 \pm 0.25$ & 0.92 & 0.95\\
\hline
\end{tabular}
\caption{Summary of constraints on the B-mode cross-correlation, $\kkB$. The table shows best-fit cross-correlation amplitude, $A_{\rm B}$, with $1 \sigma$ errors and minimum $\chi^{2}/dof$ as well as $\chi^{2}/dof$ for $A_{\rm B}=0$ (where $dof = 15$). Results are shown for cross-correlations between DES GWL from the \ngm\ catalogue and CMBWL from both SPT and \emph{Planck}. The redshift selection, $0.3 < z < 1.3$, is performed using the ${\tt SkyNet2}$ photo-$z$ catalogue.}
\label{table:results_Bmodes}
\end{table}

\subsection{Intrinsic Alignments}\label{sec:sys_IAs}

As described in \Cref{sec:theory}, GWL measurements are contaminated by the alignment of unlensed galaxy shapes determined by large-scale gravitational potentials during galaxy formation, known as intrinsic alignments (IA) \citep{HRH2000,croft00,CKB01,CNP+01,HS04}. The IAs reduce the overall observed correlation because they tend to align galaxy shapes and the matter distribution with the opposite sign to the GWL alignment.

\begin{figure}
\includegraphics[width=0.48\textwidth]{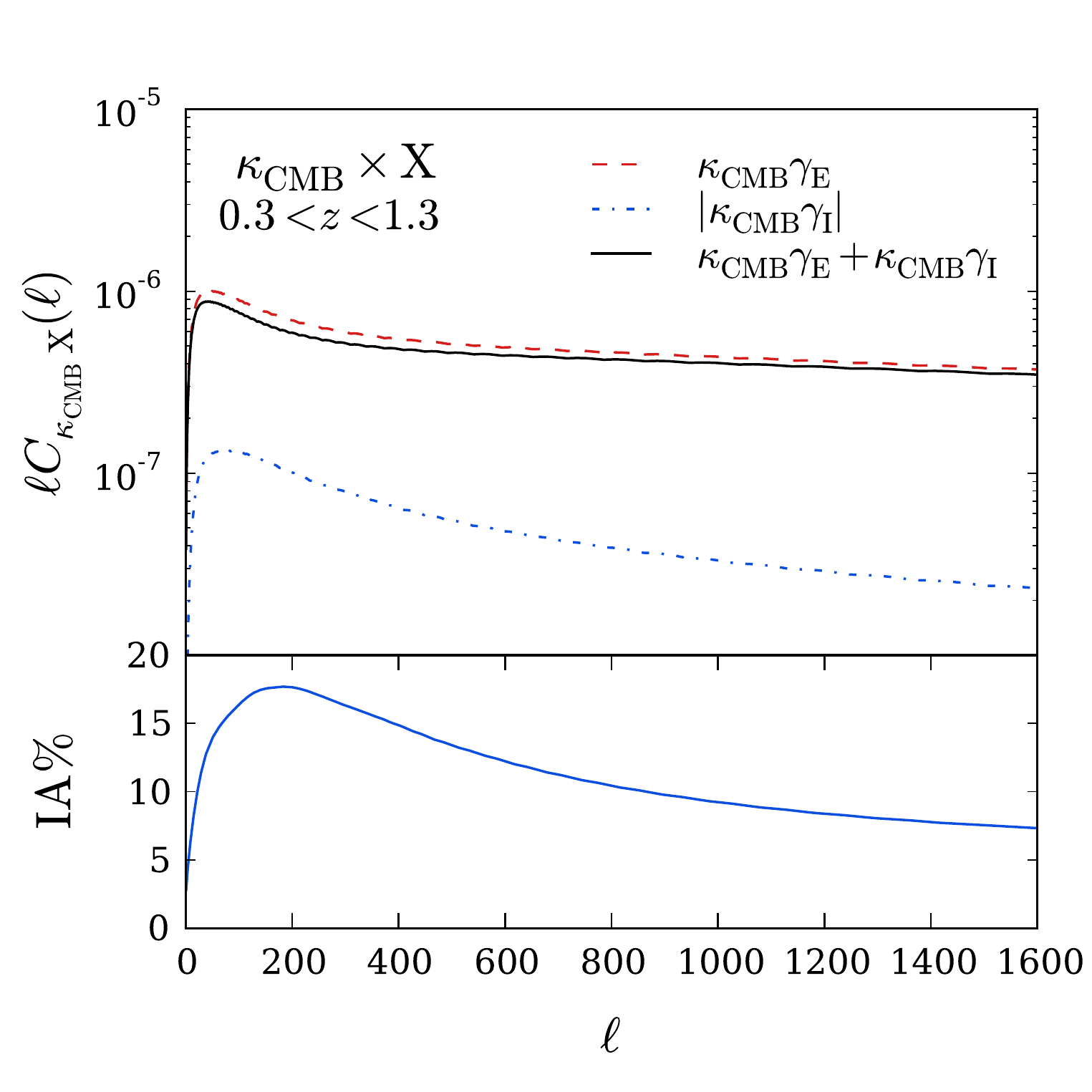}
\caption{Illustration of the impact of intrinsic alignments on our cross-correlation signal. \textbf{Top panel:} Forecast projected angular power spectra, $C(\ell)$s, for the pure GWL$\times$CMBWL cross-correlation, $\kkE$ (red dashed); the absolute value of the IA contribution to the cross-correlation, $\kkI$ (blue dot-dashed), and the total cross-correlation, $\kkE + \kkI$ (black solid). This assumes that IAs are well described by the NLA model and that they affect all galaxies equally; see \Cref{sec:sys_IAs} for more discussion. \textbf{Bottom panel:} Percentage contribution of IAs as a proportion of the total observed cross-correlation signal $|\kappa_{\rm CMB}\gamma_{I}|/(\kappa_{\rm CMB}\gamma_{E}+\kappa_{\rm CMB}\gamma_{\rm I})$. All $C(\ell)$s are calculated using CAMB sources using the ${\tt SkyNet2}$ DES SV source galaxy redshift distribution from our main analysis.}
\label{fig:sys_IAs}
\end{figure}

We examine the potential contamination of our GWL$\times$CMBWL cross-correlation by IAs using the NLA model as given in \Cref{eqn:IA_window}.
\Cref{fig:sys_IAs} shows the effect of IAs on the cross power spectrum. 
For our fiducial \texttt{SkyNet2} source redshift distribution the decrement in expected observed power spectrum, $C^{\rm obs}_{\kappa_{\rm CMB}\gamma_{\rm E}}(\ell)$, can be as much as ${\sim}$18\% around $\ell {\sim} 200$ but is substantially lower at other scales. 

This is a relatively simplistic approach to the modelling of IAs. We have assumed that the NLA model applies to all types of galaxy even though it is only designed to explain alignment of elliptical galaxies and there is, as yet, no positive detection of IAs in spiral galaxies for large surveys in the field \citep{MBB+11}. Furthermore we have assumed that the amplitude of alignment is set exactly by the $C_{1}$ normalisation of \citet{brown_spercosmos_2002,BK07}, though the DES SV cosmic shear analysis was equally consistent, at the 2$\sigma$ level, with there being no IAs or IAs at four times this assumed amplitude \citep{deswl2ptcosmo2015}. 

Our aim is to determine the significance of IAs in the detection of the cross-correlation and check that our measurement is robust to IAs.  
We defer attempts to make precise measurements of the IA signal to later work when the volume and quality of DES and SPT data will have greatly increased, producing a significant improvement in precision; see \Cref{fig:forecast} and discussion in \Cref{sec:forecasts} for more details. 

Assuming all galaxies are affected by the NLA model, with normalisation given by $C_{1}$, is a conservative way to model a `significant' IA effect that can be compared to our main analysis where IAs are ignored entirely. 
\citet{CDM+15} have recently made more careful models of IA contamination in the context of the \citet{HLD+13} data sets. Their estimated potential levels of contamination for both red and blue galaxies are slightly lower than our assumed model (${\sim} 10\%$ vs. ${\sim} 18\%$), confirming that our implementation is a realistic, conservative example.


Including IAs in this way shifts our best-fit cross-correlation amplitude from $A_{\rm no IA}=0.88\pm0.30$ to $A_{\rm with IA}=1.08\pm0.36$. The effect of IAs reduces the expected cross-correlation signal and accounting for this effect increases our measured cross-correlation amplitude. The significance of the detection remains unchanged at ${\sim} 3\sigma$ and the result is entirely consistent with forecast expectations and the measurement without IAs within the $1\sigma$ errors. Nevertheless the shift due to intrinsic alignment is at $\sim\!0.6\sigma$ level, indicating that future, higher precision measurements of this cross-correlation have the potential to be a powerful probe of IAs.  

\subsection{Photometric Redshift Uncertainties}\label{sec:sys_photoz}

Our main analyses are based on the \texttt{SkyNet2} photo-$z$ catalog but we also cross-check our main results with three other photo-$z$ estimation pipelines validated by the DES collaboration: \texttt{BPZ2}, \texttt{ANNz2} and \texttt{TPZ}. See \Cref{sec:data_des_photoz} and \citet{Bonnett2015} for more discussion of these estimators.

We recalculate our measurement of the cross-power spectrum in exactly the same way as our main analysis, but using the different photo-$z$ estimates. The photo-$z$ estimates are used to select the galaxies which enter our maps and to create the $n(z)$, which is in turn used to produce the theory $C_{\kappa_{\rm CMB}\gamma_{\rm E}}(\ell)$ used to fit the cross-correlation amplitude, $A$. See \citet{bonnettetal2015a} for more detail on the testing of these photo-$z$ pipelines in the context of the DES SV GWL analysis.


\begin{figure}
\includegraphics[width=0.48\textwidth]{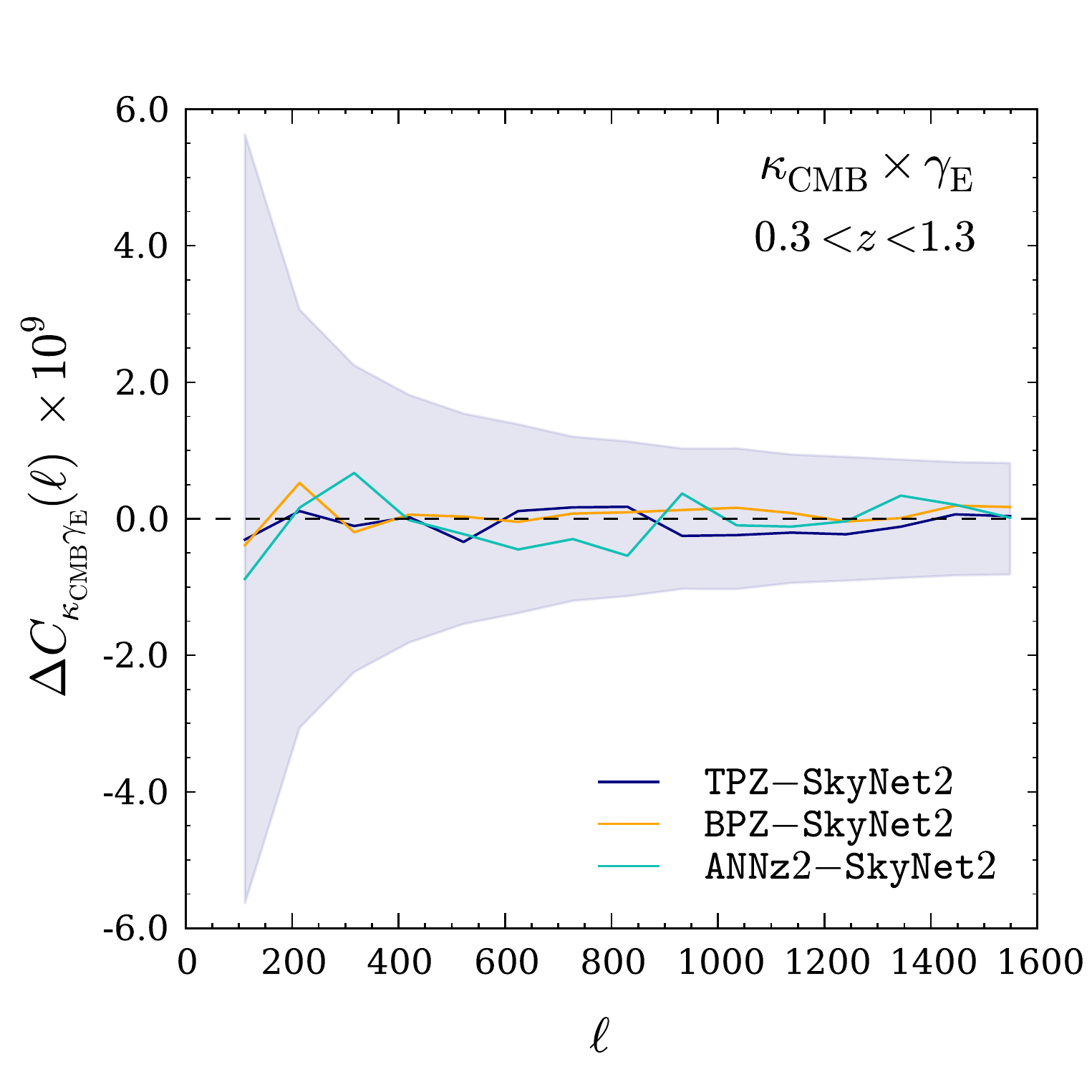}
\caption{Difference in measured power spectra, $C(\ell)$, for different choices of photometric redshift estimation pipeline, relative to the \texttt{SkyNet2} estimates used in our main analysis. All other data choices and estimator settings remain constant between runs. The grey contours show the $1\sigma$ errors on our fiducial measurement.} 
\label{fig:sys_photoz}
\end{figure}


\Cref{fig:sys_photoz} shows the difference in estimated $C(\ell)$ for each choice of estimator, relative to the fiducial \texttt{SkyNet2}. 
It is clear that the variation in measured cross-correlation due to different photo-$z$ estimation codes is significantly smaller than the error on the same measurement. 
The scatter in best-fit $A$ is well within the $1\sigma$ confidence limits. 

\citet{bonnettetal2015a} used their analysis of multiple photo-z pipelines to define a Gaussian prior on the mean of the photo-z distribution (of tomographic bins) of width $\Delta z = 0.05$. Any variation in the mean of the photo-z distribution would scale both our best-fit value of $A$ and our errors, leaving the significance of our detection unchanged. We have checked that the shift in best-fit cross-correlation amplitude due to a change of $\Delta z = \pm 0.05$ in the mean of the photo-z distribution is well within our one sigma errors and we quote the `bare' constraint on $A$ as our headline result, where we have assumed the accuracy of the \texttt{SkyNet2} $n(z)$ derived by stacking the $p(z)$ of individual galaxies.

\subsection{Systematic uncertainties in the $\kappa$ map}\label{sec:sys_kappa}
We have also tested for the degree of contamination in the SPT $\kappa$-map due to point-sources and the tSZ effect by applying a more stringent mask than the one used for lensing reconstruction.  The more stringent mask removes point sources detected between $5\sigma$ (corresponding to approximately 6 mJy) and $15\sigma$ using a $2'$ radius circular aperture in addition to the $16'\times16'$ mask applied to sources detected above $15\sigma$ using the main mask. Clusters catalogued in \citet{bleem2015} with detections between $4.5\sigma$ and $6\sigma$ are also masked with a $5'$ radius disk in addition to the  $16'\times16'$ mask applied to clusters detected above $6\sigma$. We obtain an amplitude of $A = 0.88 \pm 0.3$ with $\chi^{2}/dof = 0.98$ when applying this mask, which is entirely consistent with our main result, suggesting that our kappa maps are minimally contaminated by these sources. 

\section{Forecasts}\label{sec:forecasts}

\begin{figure}
\includegraphics[width=0.48\textwidth]{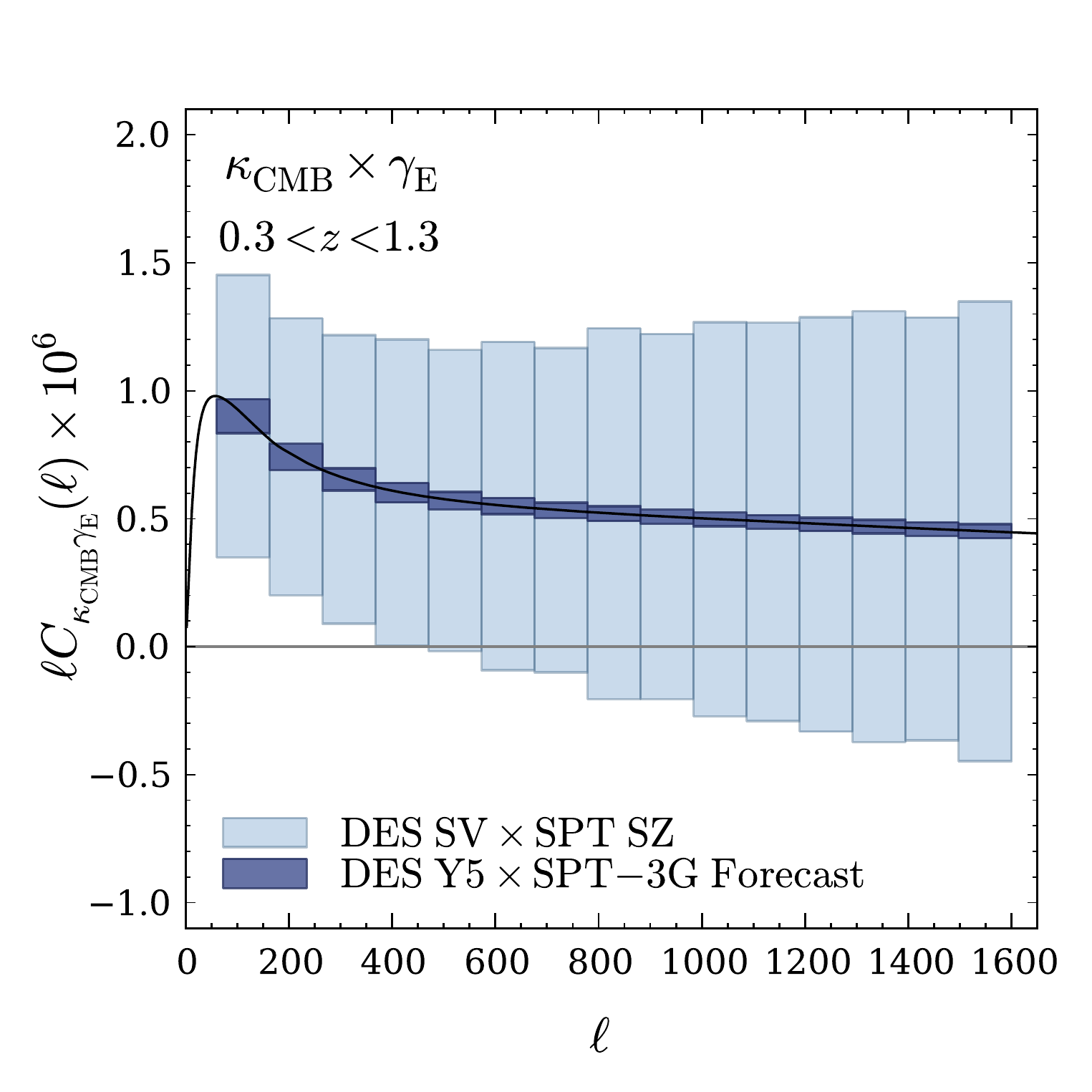}
\caption{Forecast for the DES\ Y5 GWL data cross-correlated with SPT-3G CMBWL (reconstructed from temperature plus polarisation). Shown for comparison is an analytic estimate of DES\ SV GWL cross-correlated with SPT\ SZ CMBWL. The Y5/SPT-3G (SV/SPT\ SZ) forecast assumes a sky fraction of $2500\ {\rm deg}^{2}$ ($139\ {\rm deg}^{2}$), GWL source number density $10.0\ {\rm arcmin}^{-2}$ ($5.7\ {\rm arcmin}^{-2}$) and GWL shape noise of 0.30 (0.37).} 
\label{fig:forecast}
\end{figure}

The volume of GWL surveys will greatly increase over the coming years. Dark Energy Survey\footnote{http://www.darkenergysurvey.org/} (DES) will deliver an unprecedented $5000\ {\rm deg}^2$ of lensing data by 2018, with projects including Hyper Suprime Cam (HSC)\footnote{http://www.naoj.org/Projects/HSC/}, Kilo-Degree Survey (KiDS)\footnote{http://kids.strw.leidenuniv.nl/}, \emph{Euclid}\footnote{http://www.euclid-ec.org/} and Large Synoptic Survey Telescope (LSST)\footnote{ http://www.lsst.org/} also producing data over the next decade. These surveys will push deeper than previous efforts, increasing the overlap with the CMB lensing kernel, which is broad and peaks at $z {\sim} 2$ \citep{LC06}. On the CMB side, \emph{Planck}\footnote{http://www.cosmos.esa.int/web/planck} has set a new standard for all-sky CMB surveys from space but it will also be important to maximise the overlap of galaxy surveys with high-resolution CMB surveys using the upgraded cameras on the South Pole Telescope (SPT)\footnote{https://pole.uchicago.edu/} and the Atacama Cosmology Telescope (ACT)\footnote{http://www.princeton.edu/act/}, as well as the next-generation PolarBear\footnote{http://bolo.berkeley.edu/polarbear/} instrument.

This paper represents an important test of the WL measurement pipelines in both the DES and SPT collaborations, and allows us to look forward with confidence to more scientifically ambitious analyses in the future when these more powerful data sets become available, particularly the full DES survey and the SPT third generation camera (SPT-3G) \citep{BAA+14_SPT3G}. The principal gain will be increased sky coverage, with $2500\ {\rm deg}^2$ of overlapping area expected from the full DES five year survey (Y5) and SPT-3G. This represents an ${\sim}18$-fold increase over the data used in this work. In addition, the SPT-3G upgrade will significantly decrease the noise level compared to current SPT measurements (SPT SZ). Estimates from the SPT collaboration foresee a factor of ${\sim}30$ decrease in effective noise between SPT SZ and SPT-3G when temperature measurements alone are used to reconstruct the CMBWL convergence map, and a factor of ${\sim}150$ between SPT SZ and SPT-3G when the SPT-3G reconstruction also uses CMB polarisation measurements.

\Cref{fig:forecast} shows the expected signal-to-noise (S/N) from DES Y5 and SPT-3G, compared to that from the DES SV and SPT SZ data used in this paper. We have restricted this forecast to the expected $2500\ {\rm deg}^2$ overlapping area available by DES Y5 and assumed moderate improvements in number density and GWL shape noise for DES (see figure caption for details). We can confidently expect a detection of GWL$\times$CMBWL from DES Y5 $\times$ SPT-3G with a S/N of $>50\sigma$. This huge increase in measurement power over the coming years will allow us to move beyond detection of the cross-correlation and to exploit this measurement to answer a number of science questions. Note that there is a turnover in the cross-correlation power spectrum at low ell. We have excluded this turnover from these forecasts by retaining a minimum scale of $\ell>40$. Increased coverage of this feature would further improve the power of this particular cross-correlation.

The very different observational properties of the two surveys means that the cross-correlation is an extremely useful discriminant of measurement systematics.
Both CMBWL and GWL are affected by multiplicative biases in the measurement of the lensing signal. For example, uncertainties in measuring galaxy shapes leads to a shear measurement bias in DES GWL, currently marginalised over in the cosmology analysis \citep{deswl2ptcosmo2015}.  As both probes are estimated from different types of data using very different techniques, there is considerable scope for calibration of these bias terms through cross-correlation of the GWL and CMBWL signals.

As the precision of our cross-correlation increases, the systematic effects will become more significant. In this work we estimated the order of magnitude effect of galaxy IAs, finding that the presence of IAs could shift our best-fit measurement of $A$ by a significant fraction of the $1\sigma$ errors. However, this did not alter our level of agreement with theory, given the size of our error bars. The much higher S/N measurement we can expect from future data means that the impact of IAs will be much more significant. On the one hand this means that we need to improve our modelling of IAs, paying particular attention to the impact of galaxy type and luminosity; at the same time, it means that the GWL$\times$CMBWL cross-correlation has the potential to make precision measurements of the IA signal in exactly the data sets we want to use for cosmic shear analyses, possibly allowing us to discriminate between competing IA models. 

Using this cross-correlation to measure the amplitude of the IA signal for different types of source galaxy is a real possibility. This will be important for cosmic shear studies and the first robust test of the modelling done in \citet{TI14,HT14,CDM+15} and \citet{LC15}, who have suggested that the type of spiral galaxies that dominate WL data sets may experience an IA$\times$CMBWL correlation of opposite sign to that expected in the linear alignment model. 

\section{Summary}\label{sec:discussion}

We have found evidence for the cross-correlation of GWL measured in 139 deg$^{2}$ of DES SV galaxy shape catalogues (effective source number density 5.7 arcmin$^{-2}$) and CMBWL from both SPT and \emph{Planck} at a significance of $2.9\sigma$ and 2.2$\sigma$ respectively. 
When we fit an amplitude, $A$, as a free parameter to the DES$\times$SPT cross-correlation, we measure $A = 0.88 \pm 0.30$ (68\% confidence limit), using the \emph{Planck} 2015 best-fit cosmology to calculate the expected power spectrum. The cross-correlation amplitude for DES$\times$\emph{Planck} is $A = 0.86 \pm 0.39$. Therefore we can conclude that our measurement is consistent with the expected level of cross-correlation. 


Two previous works reported detections of the GWL-CMBWL cross-correlation. \citet{HLD+13} used 121 deg$^{2}$ of Canada-France-Hawaii Telescope (CFHT) Stripe 82 GWL (12.3 source galaxies per arcmin$^{2}$) and ACT CMBWL to measure the cross-correlation with $4.2\sigma$ confidence and a best-fit amplitude of $A = 0.78 \pm 0.18$ using best-fit 2013 \emph{Planck}+lensing+WP+high-ell cosmology and $A = 0.92 \pm 0.22$ using a WMAP9 best-fit cosmology i.e. a result consistent with either the WMAP9 or \emph{Planck} expectation. In contrast, \citet{LH15} found significantly lower best-fit amplitudes for their cross-correlation of 140 deg$^{2}$ of CFHTLenS GWL (12.5 source galaxies per arcmin$^{2}$) and \emph{Planck} CMBWL. They report $A_{2013} = 0.48 \pm 0.26$ and $A_{2015} = 0.44 \pm 0.22$ using the 2013 and 2015 \emph{Planck} releases respectively. The authors speculate about a number of potential systematic effects, including IAs and photo-$z$ errors, which could produce the unexpectedly low cross-correlation amplitude. None seemed sufficient to account for the observed discrepancy. 

Our results are consistent with expectations from the \emph{Planck} 2015 cosmology and with \citet{HLD+13}, though our errors are larger due to the smaller number density of sources in the DES SV catalogues. We see no evidence of the significantly low cross-correlation amplitudes reported by \citet{LH15}, though our result is also consistent with their measurement as the $1\sigma$ errors overlap, even with their low best-fit amplitude. This is true for both our cross-correlation of DES with SPT and with \emph{Planck}.  

With three measurements of this cross-correlation now existing in the literature, there is not yet reliable evidence for any deviation from the expected LCDM level of cross-correlation, given the size of the statistical uncertainties and the significant impact of systematic errors, particularly IAs. The low best-fit amplitude found by \citet{LH15} seems to be an outlier. We have demonstrated that IAs can shift the expected result by significant fractions of the $1\sigma$ errors of the current experiments. This underlines the fact that accurate modelling of IAs \citep{TI14,HT14,CDM+15,LC15} must be an immediate priority as new data increases the precision of this measurement in the coming years.


We tested the various tools and procedures used to produce our measurement for systematic variations that could bias the result. This included checks on our measurement estimator, noise calculation and the re-making of the measurement using a different DES shape catalogue. Results when alternate estimators for both the cross-correlation and the covariance were substituted into our analysis pipeline were found to be consistent with our primary measured result. We have also tested two independent shear measurement pipelines and four independent photometric redshift estimators produced by the DES collaboration. Results from each were entirely consistent with our fiducial analysis choices. No change in the analysis procedure had a significant impact on our measured cross-correlation strength, within the errors.





CMBWL also presents us with physical phenomena that could impact future cross-correlation measurements. One example is the bias on the estimated CMB convergence arising from galaxies and clusters. These non-Gaussian foreground objects induce mode coupling of the observed sky in a way that mimics gravitational lensing. \citet{AVW+04} estimated that this contamination could be as high as 10\% based on large-scale simulations. \citet{vanengelen2014} show that biases arise from galaxy-lensing correlation and cluster-lensing correlation. The exact size of the effect depends on the choice of mask radius and maximal temperature multipole. This level of contamination does not impact our current results greatly, given the magnitude of our uncertainties, so we left it untreated in the current analysis. It will become increasingly important as we move on to DES year one data and beyond, meaning accurate estimation of the strength of this effect should be a priority.

Thinking beyond the calibration of systematic effects, as a cosmological tool GWL-CMBWL cross-correlation will be particularly important as an ingredient in a joint analysis framework with tomographic GWL and large-scale structure (LSS) data sets. Here it will help constrain cosmological parameters and calibrate systematic nuisance parameters, including galaxy bias in LSS measurements, while acting as an additional high redshift `bin'. This will increase our sensitivity to phenomena at the upper end of the redshift range of late-time probes. 
As the power of the combined data sets increases over the coming years, novel techniques to optimise cosmological information should be considered, including optimal weighting schemes to maximise the overlap of the respective sensitivity kernels of GWL, CMBWL and LSS surveys.




\section*{Acknowledgements}

We are grateful for the extraordinary contributions of our CTIO colleagues and the DECam Construction, Commissioning and Science Verification
teams in achieving the excellent instrument and telescope conditions that have made this work possible.  The success of this project also 
relies critically on the expertise and dedication of the DES Data Management group.

Funding for the DES Projects has been provided by the U.S. Department of Energy, the U.S. National Science Foundation, the Ministry of Science and Education of Spain, 
the Science and Technology Facilities Council of the United Kingdom, the Higher Education Funding Council for England, the National Center for Supercomputing 
Applications at the University of Illinois at Urbana-Champaign, the Kavli Institute of Cosmological Physics at the University of Chicago, 
the Center for Cosmology and Astro-Particle Physics at the Ohio State University,
the Mitchell Institute for Fundamental Physics and Astronomy at Texas A\&M University, Financiadora de Estudos e Projetos, 
Funda{\c c}{\~a}o Carlos Chagas Filho de Amparo {\`a} Pesquisa do Estado do Rio de Janeiro, Conselho Nacional de Desenvolvimento Cient{\'i}fico e Tecnol{\'o}gico and 
the Minist{\'e}rio da Ci{\^e}ncia, Tecnologia e Inova{\c c}{\~a}o, the Deutsche Forschungsgemeinschaft and the Collaborating Institutions in the Dark Energy Survey. 

The Collaborating Institutions are Argonne National Laboratory, the University of California at Santa Cruz, the University of Cambridge, Centro de Investigaciones Energ{\'e}ticas, 
Medioambientales y Tecnol{\'o}gicas-Madrid, the University of Chicago, University College London, the DES-Brazil Consortium, the University of Edinburgh, 
the Eidgen{\"o}ssische Technische Hochschule (ETH) Z{\"u}rich, 
Fermi National Accelerator Laboratory, the University of Illinois at Urbana-Champaign, the Institut de Ci{\`e}ncies de l'Espai (IEEC/CSIC), 
the Institut de F{\'i}sica d'Altes Energies, Lawrence Berkeley National Laboratory, the Ludwig-Maximilians Universit{\"a}t M{\"u}nchen and the associated Excellence Cluster Universe, 
the University of Michigan, the National Optical Astronomy Observatory, the University of Nottingham, The Ohio State University, the University of Pennsylvania, the University of Portsmouth, 
SLAC National Accelerator Laboratory, Stanford University, the University of Sussex, and Texas A\&M University.

The DES data management system is supported by the National Science Foundation under Grant Number AST-1138766.
The DES participants from Spanish institutions are partially supported by MINECO under grants AYA2012-39559, ESP2013-48274, FPA2013-47986, and Centro de Excelencia Severo Ochoa SEV-2012-0234.
Research leading to these results has received funding from the European Research Council under the European Union's Seventh Framework Programme (FP7/2007-2013) including ERC grant agreements 
 240672, 291329, and 306478.

This research used resources of the Calcul Quebec computing consortium, part of the Compute Canada network.

The South Pole Telescope program is supported by the National Science Foundation through grant PLR-1248097. Partial support is also provided by the NSF Physics Frontier Center grant PHY-0114422 to the Kavli Institute of Cosmological Physics at the University of Chicago, the Kavli Foundation, and the Gordon and Betty Moore Foundation through Grant GBMF\#947 to the University of Chicago.

OL acknowledges support from a European Research Council Advanced Grant FP7/291329, which also supported DK.

ABL thanks CNES for financial support through its post-doctoral programme. 

PL is funded jointly by the Royal Society of New Zealand Rutherford Foundation Trust and the Cambridge Commonwealth Trust.

CR acknowledges support from the University of Melbourne.

ES is supported by DOE grant DE-AC02-98CH10886.


\bibliographystyle{mnras}
\bibliography{refs}

\bsp


~
\newline
$[1]$ Astrophysics Group, Department of Physics and Astronomy, University College London, 132 Hampstead Road, London, NW1 2PS, United Kingdom \\ 
$[2]$ Department of Physics, McGill University, 3600 rue University, Montreal, QC, H3A 2T8, Canada \\ 
$[3]$ Sorbonne Universit\'es, UPMC Univ Paris 6 et CNRS, UMR 7095, Institut d'Astrophysique de Paris, 98 bis bd Arago, 75014 Paris, France \\
$[4]$ Fermi National Accelerator Laboratory, Batavia, IL 60510-0500, USA \\ 
$[5]$ Kavli Institute for Cosmological Physics, University of Chicago, 933 East 56th Street, Chicago, IL 60637, USA \\ 
$[6]$ Department of Astronomy and Astrophysics, University of Chicago, Chicago, IL 60637, USA \\
$[7]$ Department of Physics, ETH Zurich, Wolfgang-Pauli- Strasse 16, CH-8093 Zurich, Switzerland \\ 
$[8]$ Kavli Institute for Cosmology, University of Cambridge, Madingley Road, Cambridge CB3 0HA, United Kingdom \\ 
$[9]$ Institute of Astronomy, University of Cambridge, Madingley Road, Cambridge CB3 0HA, United Kingdom \\ 
$[10]$ Institute of Cosmology \& Gravitation, University of Portsmouth, Portsmouth, PO1 3FX, UK \\
$[11]$ Department of Physics, University of Chicago, 5640 South Ellis Avenue, Chicago, IL, USA 60637 \\
$[12]$ Institut de Ci\`encies de l’Espai, IEEC-CSIC, Campus UAB, Carrer de Can Magrans, s/n, 08193 Bellaterra, Barcelona, Spain \\
$[13]$ Centre for Theoretical Cosmology, DAMTP, University of Cambridge, CB3 0WA, United Kingdom \\
$[14]$ Department of Physics and Astronomy, University of Pennsylvania, Philadelphia, PA 19104, USA \\
$[15]$ Jodrell Bank Center for Astrophysics, School of Physics and Astronomy, University of Manchester, Oxford Road, Manchester, M13 9PL, UK \\ 
$[16]$ Department of Physics, Stanford University, 382 Via Pueblo Mall, Stanford, CA 94305, USA \\  
$[17]$ Department of Astronomy and Department of Physics, University of Illinois, 1002 West Green St., Urbana, IL 61801\\
$[18]$ Argonne National Laboratory, High-Energy Physics Division, 9700 S. Cass Avenue, Argonne, IL, USA 60439 \\  
$[19]$ Cerro Tololo Inter-American Observatory, National Optical Astronomy Observatory, Casilla 603, La Serena, Chile \\
$[20]$ Department of Physics and Electronics, Rhodes University, PO Box 94, Grahamstown, 6140, South Africa \\
$[21]$ Kavli Institute for Particle Astrophysics \& Cosmology, P. O. Box 2450, Stanford University, Stanford, CA 94305, USA \\ 
$[22]$ Carnegie Observatories, 813 Santa Barbara St., Pasadena, CA 91101, USA \\
$[23]$ Institut de F\'isica d'Altes Energies (IFAE), The Barcelona Institute of Science and Technology, Campus UAB, 08193 Bellaterra (Barcelona) Spain \\
$[24]$ SLAC National Accelerator Laboratory, Menlo Park, CA 94025, USA \\
$[25]$ Laborat\'orio Interinstitucional de e-Astronomia - LIneA, Rua Gal. Jos\'e Cristino 77, Rio de Janeiro, RJ - 20921-400, Brazil \\
$[26]$ Observat\'orio Nacional, Rua Gal. Jos\'e Cristino 77, Rio de Janeiro, RJ - 20921-400, Brazil \\
$[27]$ National Center for Supercomputing Applications, 1205 West Clark St., Urbana, IL 61801, USA \\
$[28]$ School of Physics and Astronomy, University of Southampton,  Southampton, SO17 1BJ, UK \\
$[29]$ Excellence Cluster Universe, Boltzmannstr. 2, D-85748 Garching bei M\"unchen, Germany \\
$[30]$ Universit\"ats-Sternwarte, Fakult\"at f\"ur Physik, Ludwig-Maximilians Universit\"at M\"unchen, Scheinerstr. 1, 81679 M\"unchen, Germany\\
$[31]$ Jet Propulsion Laboratory, California Institute of Technology, 4800 Oak Grove Dr., Pasadena, CA 91109, USA \\
$[32]$ Department of Astronomy, University of Michigan, Ann Arbor, MI 48109, USA \\
$[33]$ Department of Physics, University of Michigan, Ann Arbor, MI 48109, USA \\
$[34]$ Department of Astronomy, University of California, Berkeley,  501 Campbell Hall, Berkeley, CA 94720, USA \\
$[35]$ Lawrence Berkeley National Laboratory, 1 Cyclotron Road, Berkeley, CA 94720, USA \\
$[36]$ Max Planck Institute for Extraterrestrial Physics, Giessenbachstrasse, 85748 Garching, Germany \\
$[37]$ Center for Cosmology and Astro-Particle Physics, The Ohio State University, Columbus, OH 43210, USA \\
$[38]$ Department of Physics, The Ohio State University, Columbus, OH 43210, USA \\
$[39]$ Australian Astronomical Observatory, North Ryde, NSW 2113, Australia \\
$[40]$ Departamento de F\'isica Matem\'atica,  Instituto de F\'{\i}sica, Universidade de S\~ao Paulo,  CP 66318, CEP 05314-970, S\~ao Paulo, SP,  Brazil \\
$[41]$ Department of Astronomy, The Ohio State University, Columbus, OH 43210, USA \\
$[42]$ Department of Astrophysical Sciences, Princeton University, Peyton Hall, Princeton, NJ 08544, USA \\
$[43]$ Instituci\'o Catalana de Recerca i Estudis Avan\c{c}ats, E-08010 Barcelona, Spain \\
$[44]$ School of Physics, University of Melbourne, Parkville, VIC 3010, Australia\\
$[45]$ Department of Physics, University of Arizona, Tucson, AZ 85721, USA \\
$[46]$ Centro de Investigaciones Energ\'eticas, Medioambientales y Tecnol\'ogicas (CIEMAT), Madrid, Spain \\
$[47]$ Instituto de F\'isica Te\'orica, Universidade Estadual Paulista, Rua Dr. Bento T. Ferraz 271, Sao Paulo, SP 01140-070, Brazil \\
\label{lastpage}

\end{document}